\documentclass[11pt,nonatbib]{elsarticle}
\usepackage{eurosym}
\usepackage{amssymb}
\usepackage{graphics,graphicx,endnotes,amsmath,amsfonts,bm}
\usepackage{indentfirst}
\usepackage{multirow}
\usepackage{pdflscape}
\usepackage{array, threeparttable, booktabs, caption}
\usepackage{adjustbox}
\usepackage{bm}
\usepackage{geometry}
\usepackage{ragged2e}
\usepackage{appendix}
\usepackage{csquotes}

\usepackage[english]{babel}

\let\origappendix\appendix
\renewcommand\appendix{\clearpage\pagenumbering{roman}\origappendix}
\renewcommand{\thetable}{\arabic{table}}

\renewcommand{\thefootnote}{\fnsymbol{footnote}}
 \makeatletter
\def\@biblabel#1{\hspace*{-\labelsep}}
\makeatother
\def\code#1{\texttt{#1}}

\usepackage[colorlinks=true,linkcolor=blue,urlcolor=blue,citecolor=blue]{hyperref}

\begin{document}

\thispagestyle{empty} \noindent \bigskip

\begin{center}
\textbf{\LARGE Priced risk in corporate bonds}{\large \footnote{%
Nikolai Roussanov was the editor for this article. We gratefully acknowledge comments and suggestions from an anonymous referee, Patrick Augustin, Kevin Aretz, Mikhail Chernov, Francis Cong, Mathieu Fournier, Patrick Houweling, Yoshio Nozawa, Piotr Or{\l}owski, Nikolai Roussanov, Avanidhar Subrahmanyam, and seminar and workshop participants at the Board of Governors of the Federal Reserve System, Erasmus University Rotterdam,  ESSEC Business School, Luiss University, Norwich Business School, and Robeco Asset Management. The companion website to this paper, \href{https://openbondassetpricing.com/}{\texttt{openbondassetpricing.com}}, contains full replication code and updated corporate bond factor data.
}}
\end{center}

\leavevmode

\begin{center}
{\large Alexander Dickerson}$^{\text{a}}${\large , Philippe Mueller}$^{\text{b}}$, {\large and Cesare Robotti}$^{\text{%
c,}}${\large \footnote{%
Corresponding author.
\par
\ \textit{Email addresses}: \texttt{alexander.dickerson1@unsw.edu.au}
(A.~Dickerson), \texttt{philippe.mueller@wbs.ac.uk}
(P.~Mueller), and \texttt{cesare.robotti@wbs.ac.uk} (C.~Robotti).}}
\end{center}

\begin{center}
\noindent $^{\text{a}}$\textit{UNSW Business School, Sydney, NSW 2052, Australia} \\ $^{\text{b,c}}$\textit{Warwick Business School, Coventry, CV4 7AL, United Kingdom}\bigskip \bigskip
\end{center}

\begin{center}
August 2023
\end{center}

\leavevmode

\noindent \text{{\large ABSTRACT}}

\noindent Recent studies document strong empirical support for multifactor models that aim to explain the cross-sectional
variation  in corporate bond expected excess returns.
 We revisit these findings and provide evidence that common factor pricing in corporate bonds is exceedingly difficult to establish. Based on portfolio- and bond-level analyses, we demonstrate that previously proposed bond risk factors, with traded liquidity as the only marginal exception, do not have any incremental explanatory power over the corporate bond market factor.
Consequently, this implies that the bond CAPM is not dominated by either traded- or nontraded-factor models in pairwise and multiple model comparison tests.

\vspace*{0.5in} \noindent \textit{Keywords}: Corporate bond pricing; Bond CAPM; Sharpe ratio; Efficient
frontier; Model misspecification and identification.

\vspace*{0.5in} \noindent \textit{JEL classification}: C12; C13; G12.

\newpage \setcounter{page}{1} \renewcommand{\thefootnote}{\arabic{footnote}}
\setcounter{footnote}{0}

\newpage \setcounter{page}{1}\parskip0.5em \baselineskip 20pt

\section{Introduction}
Over the past fifty years, significant strides have been made  in the understanding of the determinants of the pricing kernel for equities. The cornerstone of the literature that relates to identifying a common factor structure in expected returns is the Capital Asset Pricing Model (CAPM) with associated value-weighted stock market factor. The empirical failure of the CAPM has been extensively documented and has spurred a vast literature on additional pricing factors that seek to explain the cross-section of expected equity returns. In turn, characterization of factors that form the pricing kernel for corporate bonds has only recently been studied more comprehensively.
 For the most part, the proposed risk factors for corporate bonds differ from those for stocks and are typically derived from term structures of interest rates or metrics related to bond liquidity and default risk. Although both areas of study have mostly developed independently, recent attempts to identify a common factor structure across asset classes have only yielded mixed success.
Furthermore, as for equities, the search for theory-motivated risk factors in the cross-section of corporate bonds is still an ongoing effort.

In this paper we revisit the main findings of a series of prominent papers
on corporate bond pricing, with a particular emphasis on Bai, Bali, and Wen (2019, henceforth BBW), by utilizing well-known economic metrics and a comprehensive set of statistical tools. We show across various bond databases that outperforming a single-factor model with the bond market factor ($MKTB$) as the only source of systematic risk is a challenging task. This finding is in stark contrast to BBW and others who claim that factors beyond the market provide nontrivial incremental pricing ability for corporate bonds. For example, BBW argue that the downside ($DRF$), credit ($CRF$), and liquidity ($LRF$) risk factors are not spanned by existing factors (including $MKTB$ as well as the bond factors of Fama and French, 1993), thus enhancing the explanatory power of the model.
We reconcile these contrasting findings by showing that the publicly available $DRF,$ $CRF,$ and $LRF$ provided by BBW are not properly constructed and suffer from lead/lag errors over extended portions of their sample period. (The sample period considered by BBW is August 2004 to December 2016.) Moreover, we show that truncating both tails of $MKTB$ as in BBW, reduces its risk premium and favors alternative factors in multivariate tests.

Our main finding of a lack of incremental pricing ability for all of the newly proposed bond factors is important, as the initial asset pricing findings for corporate bonds documented by earlier studies have laid tenuous groundwork for further research.
As an example, the BBW four-factor model has become a de facto benchmark for assessing the risk-adjusted performance of traded risk factors in the corporate bond market, akin to the Fama and French (1993) three-factor model in the evaluation of new stock-related factors/anomalies.
To investigate the economic relevance of additional traded factors, beyond that of the bond market factor, we first employ widely-used metrics such as the distance from the mean-variance frontier and maximum Sharpe ratio analysis. Our exploratory analysis provides preliminary evidence that the additional factors proposed in prior work do not outperform the value-weighted bond market factor from an investment perspective. That is, the improvement to an investor's portfolio from including these factors above and beyond holding the bond market portfolio is statistically and economically marginal, at best.

Next, we employ misspecification- and identification-robust time-series and cross-sectional regression techniques to investigate whether the proposed traded factors exhibit substantial pricing ability for portfolios sorted on maturity, industry, ratings, and credit spreads. We closely follow Lewellen, Nagel, and Shanken (2010, henceforth LNS) in our analysis who argue that the ordinary least squares (OLS) estimator  provides little economic interpretation and can yield large $\text{R}^2$s  even when the fundamental asset pricing relation is violated. In a situation with substantial heterogeneity in returns across different test portfolios -- as is the case for corporate bonds -- LNS advocate the use of the generalized least squares (GLS) cross-sectional regression (CSR) $\text{R}^2$, which downweights the informational content of standard OLS-based goodness-of-fit measures. When applying GLS estimation and inference, the initial findings of some pricing ability for the non-market traded bond factors are substantially weakened. At a portfolio level, only traded liquidity ($LRF$) seems to have nontrivial incremental explanatory power over the bond market factor when using a GLS weighting scheme.

Our analysis highlights several statistical issues that are frequently overlooked in empirical asset pricing. Among them, model misspecification and uncertainty are particularly salient. As economic theory provides only limited guidance on the precise structure of the pricing kernel, it seems prudent to explicitly recognize model uncertainty associated with evaluating the pricing ability of a set of risk factors across different test asset portfolios. Model misspecification becomes a particularly prominent issue in a situation where factors are chosen ex post and merely based on empirical relevance rather than a theoretical foundation.
As all financial models are approximations of reality, they are likely to be misspecified. It is therefore imperative to incorporate this additional source of uncertainty in statistical inference. Fortunately, misspecification adjustments for both linear and nonlinear models are now readily available, and such robust inference methods remain valid even when the model is correctly specified. Employing these adjustments (with the GLS ones being also robust to a model's possible lack of identification) further reinforces the conclusion that the additional factors proposed in prior research, with the marginal exception of $LRF,$ exhibit no incremental pricing ability over the bond market factor.

In contrast to equities, bonds trade on an over-the-counter market, where trading costs are substantially higher and fluctuate significantly. (See Bao, Pan, and Wang, 2011, and Dick-Nielsen
and Rossi, 2019, among others.) Since liquidity risk has been shown to be priced in the cross-section of stock returns (Amihud, 2002, and P\'astor and Stambaugh, 2003), it would seem intuitive that the liquidity risk premium should extend to corporate bonds. Findings by Goldberg and Nozawa (2021) suggest that liquidity supply by financially constrained intermediaries is a key driver of corporate bond market liquidity and prices. Capturing the effects of illiquidity on asset prices is most often achieved via nontraded factors. Given the trading environment of corporate bonds, it would seem compelling to posit that bond exposure to liquidity shocks should yield explanatory power for the time-series and cross-section of expected bond returns.
Besides illiquidity, a few  papers  examine the asset pricing implications of corporate bond exposure to systematic volatility (as captured by the Chicago Board Options Exchange (CBOE) volatility index (VIX)), macroeconomic uncertainty (as captured by the macro uncertainty index of  Jurado, Ludvigson, and Ng, 2015), and long-run consumption risk (as in  Elkamhi, Jo, and Nozawa, 2023, henceforth EJN).

We investigate the ability of these nontraded factors in pricing the cross-section of corporate bond returns in the context of the beta-pricing models in which they were originally considered. Prior work Chung, Wang, and Wu (2019), Lin, Wang, and Wu (2011),  Bali, Subrahmanyam, and Wen (2021), and EJN reveals great empirical support for systematic volatility, liquidity, macroeconomic uncertainty, and long-run consumption risk as priced risk factors in the cross-section of corporate bond returns. In revisiting these findings, we conclude that all of the proposed nontraded factors are redundant and often spurious, that is, these factors do not command statistically significant risk premia and do not improve on the risk-return trade-off for corporate bonds.
Moreover,  the risk premia (associated with these nontraded factors) from two-pass cross-sectional regressions  are economically small and statistically insignificant.

While our findings so far are either based on factor time-series analyses or CSRs at a portfolio level, we also explore the bond-level pricing implications of the various factors/models considered. Specifically, we investigate whether the above factors are priced in the cross-section of individual bond returns based on predictive
Fama and MacBeth (1973) regressions. The results from the bond-level analysis largely confirm our time-series and portfolio-level results in the first part of the paper. Interestingly, the bond market factor ($MKTB$) as well as the traded liquidity factor ($LRF$) continue to exhibit nontrivial pricing power even at an individual bond level, while the other considered traded and nontraded factors do not seem to add to the explanatory power of the various proposed models. Our portfolio- and bond-level findings are robust across bond databases.

It is important to stress that we do not attempt to reconcile the empirical findings with an underlying theoretical foundation in this paper, but we primarily focus on empirical and methodological considerations.
While we believe that it is potentially valuable from a practical perspective to include factors that capture downside, credit, or liquidity risk, we highlight some major challenges that empirical asset pricing is facing in the context of corporate bonds, and
we provide some guidance and recommendations aimed at ensuring more dependable and less precarious inference and evaluation of asset pricing models and factors. Our objective is to provide direction for empirical asset pricing research in corporate bonds and to encourage further exploration of this area.

The paper is organised as follows.
In Section~\ref{sec:data}, we discuss the relevant data used in empirical asset pricing studies for corporate bonds. In Section~\ref{sec:common_factors}, we present descriptive evidence for the BBW four-factor model and other models with traded factors,  and we highlight important replication issues.
Sections~\ref{sec:frontier} and \ref{sec:fit}
introduce inference tools and procedures to assess the robustness of the findings on the pricing ability of the considered traded factors. Specifically, we present results for both time-series and cross-sectional tests based on portfolio-level analysis. In Section~\ref{sec:nontraded}, we examine nontraded-factor models while in Section~\ref{sec:bond_ap}, we perform bond-level analysis for models with only traded factors and for models with traded and nontraded factors, respectively. Section~\ref{sec:conclusion} summarizes our main conclusions.
Additional data work can be found in the Appendix. Finally, the analyses with alternative databases and our reply to Bai, Bali, and Wen (2023) are included in the Internet Appendix.

\section{Data and variable definitions}\label{sec:data}
This section provides a detailed discussion related to generating monthly corporate bond data using the Trade Reporting and Compliance Engine (TRACE) database and the Mergent Fixed Income Securities Database (FISD).

\subsection{Enhanced TRACE corporate bond data}
For the main results, we use the Enhanced TRACE
database for the sample period July 2002 to December 2016.
TRACE offers intraday clean bond prices, trading volumes, and buy and sell indicators for all over the counter bond trades in the U.S. We then combine the TRACE bond pricing information with FISD to acquire additional bond characteristic data.
We closely follow the prescribed bond filtering rules from the literature: i) Remove bonds that are not publicly traded in the U.S. market. These include bonds issued through private placement, bonds issued under Rule 144A, bonds that are not traded in US dollars, and bonds from issuers not based in the United States; ii) Remove bonds that are classified as structured notes, mortgage backed or asset backed, agency backed, equity linked or convertible;
iii) Remove bonds that have a floating coupon rate;
iv) Remove bonds that have less than one year remaining until maturity;
v)  Remove all intraday transactions for which the trade price is less than \$5 or greater than \$1,000. This implicitly removes many bonds that have defaulted or are close to default; vi) Eliminate all bond transactions that are labelled as when-issued, locked-in, or have special sales conditions, and that have more than a two-day settlement; vii) Remove transaction records that are cancelled and adjust records that are subsequently corrected or reversed. We follow the procedures outlined in
Dick-Nielsen (2014) to achieve this filtering;
 and viii) Remove intraday transaction records that have trading volume less than \$10,000.

\subsection{Corporate bond returns}
To generate monthly time series of corporate bond prices, and to calculate monthly bond returns, one requires the clean price of the bond ($P$) and accrued interest ($AI$) for each bond at the end of each month. We follow the literature and first calculate the clean price of each corporate bond on each day $d$ as the volume-weighted average of the intraday bond trade prices.
Thereafter, we use the daily bond prices to compute monthly returns.
Specifically, the return computation method identifies two possibilities for a valid bond return to be realized at the end of any given month $t:$ (i) there is a valid price at the end of month $t$ and the end of month $t-1;$ and (ii) there is a valid price at the beginning of month $t$ and the end of month $t.$ The `end of the month' condition is satisfied if the bond trades in the last five trading (business) days of any month $t.$ Conversely, the `beginning of the month' condition is met if the bond trades in the first five trading days of  month $t.$ This implies that if the bond price is missing during the last five trading days of month $t-1,$ the price observation is set equal to the price on the first available day of the first five business days of month $t.$  If a monthly price can be obtained in both scenarios, preference is given to the price that is recorded in the last five business days of month $t-1$ (with priority being given to the price on the last available business day of this five-day window) as opposed to the first five business days of month $t.$

The monthly corporate bond return at time $t$ for each bond $i$ in the sample is then computed as
\begin{equation}
    R_{i,t} = \frac{P_{i,t} + AI_{i,t} + C_{i,t}}{P_{i,t-1} + AI_{i,t-1}} - 1,
    \label{eq:Equation_BR}
\end{equation}
where $P_{i,t}$ is the clean bond price for bond $i$ at time $t$ as described above, $AI_{i,t}$ is the accrued interest for bond $i$ at time $t,$ and $C_{i,t}$ is the coupon (if any) for bond $i$ at month $t.$
The bond return in excess of the one-month U.S. T-Bill rate of return, $r_{f,t},$ is computed as $r_{i,t} = R_{i,t} - r_{f,t}.$
The final sample comprises 31,348 bonds issued by 3,792 firms with a total of 861,524 bond-month observations over the sample period of July 2002 to December 2016. On average, we observe 4,951 bonds per month over the sample. (In the Appendix, we delve deeper into some of the main features of our data from the TRACE and FISD databases.)

\section{Common factors in the corporate bond market}\label{sec:common_factors}
We start this section by describing the most prominent factor model for excess returns on corporate bonds. We then introduce other traded-factor models that have shown some mixed success in explaining the cross-sectional variation in corporate bond returns. Finally, we describe the set of test portfolios that we employ in the CSRs at a portfolio level.

\subsection{The BBW four-factor model}
BBW propose a four-factor model to price the cross-section of corporate bond returns. The model includes
the bond market factor $MKTB,$ the downside risk factor $DRF,$ the credit risk factor $CRF,$ and the liquidity risk factor $LRF.$  We strictly follow their paper to construct these factors. In contrast to BBW, we do not rely on winsorization of bond excess returns in factor construction.

\paragraph{The bond market factor ($MKTB$)}
The corporate bond market factor is constructed akin to the equity market factor, i.e., $MKTB$ is proxied by the average return on all bonds in our sample, weighted by their amounts outstanding.

\paragraph{The downside risk factor ($DRF$)}
For each bond, we estimate the historical 5\% Value at Risk (VaR) from monthly returns in the past 36 months (requiring a minimum of 24-months of data) and use its absolute value as the downside risk measure (VaR5).
To construct $DRF$, for each month we independently sort bonds into $5 \times 5$ portfolios according to their ratings and VaR5. Then, for each rating quintile, we calculate the weighted average return difference between the highest VaR5 quintile and the lowest VaR5 quintile. Finally, $DRF$ is computed as the average long-short portfolio return across all rating quintiles.

\paragraph{The liquidity risk factor ($LRF$)}
We first compute a bond-level proxy for illiquidity. The illiquidity measure ($ILLIQ$) is defined as
$-\text{Cov}_t (\Delta p_{i,t,d}, \Delta p_{i,t,d+1}),$ where $\Delta p_{i,t,d} \equiv \text{log}(P_{i,t,d}) - \text{log}(P_{i,t,d-1})$ is the log price change for bond \textit{i} on day \textit{d} of month \textit{t}, and the covariance is calculated over all daily returns in month \textit{t} (Bao, Pan, and Wang, 2011). To recognize a daily bond return, we require that the number of trading days between the lagged price and the current price be less than or equal to one week (7 business days). We also require at least 5 observations of the paired price changes, $(\Delta p_{i,t,d}, \Delta p_{i,t,d+1})$.\footnote{Using a different number of paired price changes or number of trading days between daily returns does not make a material difference to the $LRF$ factor mean or Sharpe ratio.}
This procedure yields an estimate of $ILLIQ$ for each bond \textit{i} in the sample at the end of month \textit{t}. To construct $LRF,$ for each month we independently sort bonds into $5 \times 5$ portfolios according to their ratings and $ILLIQ.$ Then, for each rating quintile, we calculate the weighted average return difference between the highest $ILLIQ$ quintile and the lowest $ILLIQ$ quintile. Finally, $LRF$ is computed as the average long-short portfolio return across all rating quintiles.

\paragraph{The credit risk factor ($CRF$)} $CRF$ is the equally-weighted average return on three `credit portfolios': $CRF_{\text{VaR}},$ $CRF_{ILLIQ},$ and $CRF_{REV}.$  The credit risk factor across the VaR5 portfolios, $CRF_{\text{VaR}},$ is the value-weighted average return difference between the lowest-rating (i.e., highest credit risk) portfolio and the highest-rating (i.e., lowest credit risk) portfolio across the VaR portfolios. The credit risk factor across the $ILLIQ$ portfolios, $CRF_{ILLIQ},$ is the value-weighted average return difference between the lowest-rating portfolio and the highest-rating  portfolio across the $ILLIQ$ portfolios. Finally, the credit risk factor across the reversal portfolios, $CRF_{REV},$ is the value-weighted average return difference between the lowest-rating  portfolio and the highest-rating  portfolio across quintiles sorted on $REV.$

\paragraph{Replicated vs. original BBW factors}
In Fig.~1, we plot the time series of our replicated factors and the original BBW factors, obtained from Turan Bali's personal webpage.\footnote{Shortly after the public release of our working paper in May 2023, the original TRACE-based factors from BBW have been removed from both Turan Bali's and Jennie Bai's personal websites, and they have been replaced with new factors based on the Wharton Research Data Services (WRDS) database. We provide the original BBW factors, which contain the lead-lag errors, and the correctly-constructed replicated factors for download on the companion website \href{https://openbondassetpricing.com/}{\texttt{openbondassetpricing.com}}.}

\begin{center}
\fbox{Figure~1 about here}
\end{center}
First, there is a lead error (look-ahead bias) in the original $DRF$ and $CRF$ factors for the majority of the sample. That is, the return value for month $t$ is in fact the value for month $t+1.$ This error spans the sample period from 2004-08-31 to 2014-12-31 (the entire sample except for 2015 and 2016). For $LRF$, there is a lag error which spans the sample period 2015-01-31 to 2016-12-31 (the last two years of the sample), i.e., the returns for month $t$ are the returns for month $t-1.$ Both the replicated and original $MKTB$ factors are free from these errors.
Second, it is also quite clear, across all of the original factors, that the negative return realizations are much more attenuated than for the replicated factors.
We correct the original $DRF$ and $CRF$ factors for the lead and lag errors and present summary statistics for the original and the replicated factors in Panels~A and B of Table~1 below. We present the pairwise correlations between the original and replicated factors in Panel~C.1 and between the lead-lag corrected and replicated factors in Panel~C.2. Finally, the time-series correlations between the original and the replicated factors are presented in Panels~D.1 and D2.

\begin{center}
\fbox{Table~1 about here}
\end{center}
Based on Panels~A and B, the replicated factor means are not too far from their original counterparts. However, the standard deviations of the replicated factors are much higher. The standard deviations for the replicated $DRF$ and $CRF$ are  41\% and 82\% higher than those of the original series. This confirms the finding in Fig.~1, where the negative factor returns are strongly attenuated for the original factors, which drives down the factor standard deviations. The same phenomenon occurs for $MKTB$ and its standard deviation, which is 37\% higher in the replication. Turning to the percentiles, the minimum factor values for the replication are much smaller. Across all of the factors, the minimum factor returns for the replicated factors are, besides $MKTB,$ more than double the minimum values for the original factors. For the case of $CRF,$ the minimum value in the replication is $-21.91\%$ compared to $-8.84\%$ in the original.  Interestingly, the maximum values of the replicated and original factors are much more aligned.

Turning to Panel C, the pairwise correlations between the original factors (with the lead-lag error) and the replicated factors are presented in Panel~C.1. The correlations for the $DRF$ and $CRF$ factors are very low at 26\% and 44\% respectively, presumably because, for over 80\% of the sample, the two original factors are affected by a one-month look-ahead bias.
The pairwise correlations for $MKTB$ and $LRF$ are reasonably high at 94\% and 83\%, respectively. In Panel~C.2, we report the pairwise correlations between the corrected original factors and the replicated factors. The correlations for $DRF$ and $CRF$ now jump up to over 90\%, and the correlation for $LRF$ increases to 88\%. Panel~D.1 reports the factor correlations between the original factors, and Panel~D.2 the factor correlations between the replicated factors. For example, the correlation between $MKTB$ and $DRF$ is 79\% in the replication and only 28\% for the original series. Similarly, the correlations between $LRF$ and $MKTB$ (47\% vs. 62\%) and $LRF$ and $DRF$ (32\% vs. 80\%) are also markedly higher in the replication.
For the main body of this paper and the remainder of the results, we rely on the replicated four factors using the Enhanced TRACE database.

\subsection{Additional traded-factor models}

We also analyze the pricing performance of some other models with traded factors that have been considered in corporate bond pricing.

\paragraph{The Capital Asset Pricing Model (CAPM)}
We include the value-weighted stock market excess return ($MKTS$), which is obtained from Kenneth French's webpage. The inclusion of this factor is motivated by Binsbergen, Nozawa, and Schwert (2023, henceforth BNS), who show that the equity CAPM is effective in pricing duration-adjusted corporate bond returns (bond returns that are stripped of term structure risk).\footnote{Since most empirical work on corporate bonds is concerned with returns in excess of the one-month T-Bill rate, we stick with this convention to aid comparability with prior studies. However, when using duration-adjusted returns, the main message of our paper is reinforced and confirms the findings in BNS. The results for the analysis based on duration-adjusted returns is available from the authors on request.}

\paragraph{The default and term structure model (DEFTERM)}
Default risk ($DEF$) and term structure risk ($TERM$) were introduced by Fama and French (1993).\footnote{The data to construct both factors are available on  \href{https://sites.google.com/view/agoyal145}{Amit Goyal's webpage}.}
$DEF$ is defined as the return difference between the market portfolio of long-term corporate bonds (the Composite portfolio of the corporate bond module of Ibbotson Associates) and long-term government bonds. The factor is meant to capture changes in economic conditions that affect the likelihood of default. $TERM$ should proxy for common risk in bond returns induced from unexpected changes in interest rates, and it is defined as the return difference between long-term government bonds and the one-month T-Bill rate.

\paragraph{The intermediary capital models (HKM and HKMSF)}
Intermediary asset pricing provides a novel view on the role of financial intermediaries in the pricing of major asset classes. Recent work by He, Kelly, and Manela (2017, henceforth HKM) shows that a two-factor model,  including shocks to the equity capital ratio of New York Fed's primary dealers and the stock market factor, has nontrivial explanatory power for the cross-section of returns on seven asset classes and corporate bonds in particular. Given the possibly strong pricing ability of this two-factor model, it is natural to include it in a more thorough analysis that is concerned only with corporate debt. Following HKM, we employ the value-weighted equity excess return for the New York Fed's primary dealer sector ($CPTLT$) as a proxy for traded financial intermediary risk. In addition, similar to HKM, we also consider in our analysis a single-factor model with $CPTLT$ only and denote this model by HKMSF.\footnote{We thank Zhiguo He, Bryan Kelly, and Asaf Manela for making the traded and nontraded versions of their financial intermediary factor publicly available on their websites.}

\subsection{Test portfolios}
For our cross-sectional investigations at a portfolio-level, we construct a set of bond portfolios that are sorted on various bond characteristics and the Fama-French industry classification standards. Specifically, we include 5 portfolios sorted on bond rating, 5 portfolios sorted on maturity, 10 portfolios sorted on credit spread, and the 12 Fama-French industry portfolios, so that the total number of portfolios is $N=32.$\footnote{Given the relatively short time-series sample size of $T=149,$ we try to keep the total number of portfolios as small as possible while ensuring that the cross-sectional dispersion in portfolio returns is sufficiently high. This is important because some of the tests that we employ are only asymptotically justified (i.e., they require $T \rightarrow \infty$ and $N$ to be fixed) and may not work very well in finite samples especially when $N$ is large relative to $T.$} The addition of the 12 industry portfolios follows the advice of LNS in that their returns do not necessarily relate to the aforementioned risk characteristics.
Furthermore, the inclusion of portfolios sorted on credit spreads is motivated by the work of Nozawa (2017) who finds that bond credit spreads are an important driver of the cross-sectional variation in excess corporate bond returns.
We follow the credit spread portfolio formation method in EJN and form 10 portfolios based on the average bond credit spreads between months $t-12$ and $t-1.$ The one-month lag between the signal observation and the portfolio formation month reduces measurement error in the one-month ahead realized average portfolio returns.

\section{Summary statistics, mean-variance frontiers, and Sharpe ratios}\label{sec:frontier}

In this section, we revisit the evidence provided by BBW from an economic perspective and explore the cross-sectional asset pricing implications later on.
We report the sample means for the factors comprising the four factors of BBW, $DEF,$ $TERM,$ $MKTS,$ and $CPTLT$ in Panel A of Table~2. Associated \textit{p}-values are presented in square brackets below.\footnote{Throughout the paper, unless mentioned otherwise, all the results account for conditional heteroskedasticity and for possible serial correlation in the data based on the heteroskedasticity and autocorrelation consistent estimator of Newey and West (1987, henceforth NW). Since most of the autocorrelations of the relevant terms in the expressions for the asymptotic variances of our tests are relatively small (under 0.2 and frequently under 0.1) and often not statistically significant, we apply a three-lag NW adjustment.}

\begin{center}
\fbox{Table~2 about here}
\end{center}
All of the BBW factor means are significantly different from zero at the 5\% level except for $CRF$.\footnote{Given the relatively large transaction costs associated with trading corporate bonds and the widespread evidence of model misspecification in the BBW model that we document later, the factor means should be taken with caution if we were to interpret them as risk premia.} The standard deviations of $DRF$ and $CRF$ are almost double those of $MKTB$ and $LRF,$ and an inspection of the pairwise correlation between $DRF$ and $MKTB$ in Table~1 reveals that $DRF$ is essentially a more noisy version of $MKTB.$ The $DEF$ mean is tiny in magnitude and insignificant while the $TERM$ mean is large and statistically significant at the 10\% level. The mean of the value-weighted stock market factor is also significant at the 10\% level while the $CPTLT$ mean is statistically insignificant.

Next, we employ the $MKTB$ factor to compute the one-factor alpha across all of the remaining factors. Surprisingly, after adjusting for bond market risk, the $DRF$, $CRF,$ and $LRF$ premia (now interpreted as an `alpha') drop drastically and are not statistically different from zero.\footnote{The $p$-values associated with the alphas are based on a conditional heteroskedastic version of the test of Gibbons, Ross, and Shanken (1989, henceforth GRS) as in Barillas, Kan, Robotti, and Shanken (2020, henceforth BKRS).} In fact, at the 5\% significance level, the only factor that does not appear to be spanned by the bond market factor is $DEF$ (default risk). However, its average abnormal return is negative ($-$0.31\% per month), making $DEF$ quite unattractive from an investment perspective.
Overall, these preliminary findings are in sharp contrast to the results presented in BBW, where even after using a nine-factor benchmark, the alphas on their credit, default, and liquidity factors remain alarmingly large and highly statistically significant.

To further understand the properties of the BBW factors, we examine model performance in terms of bias-adjusted squared Sharpe ratios.  We report $p$-values for the test of the null hypothesis that the true squared Sharpe ratio is equal to zero.\footnote{The estimated squared Sharpe ratio for each factor (model) is modified so as to be unbiased in small samples under normality (joint normality). Let $T$ and $K$ denote the number of time-series observations and factors, respectively. This entails multiplying the sample squared Sharpe ratio by $(T-K-2)/T$ and subtracting $K/T,$  eliminating the upward bias, while leaving
the asymptotic distribution unchanged. The details of the analysis are provided in BKRS.} Interestingly, among all factors considered, $MKTB$ yields the highest bias-adjusted squared Sharpe ratio, followed by $LRF$ and $DRF.$ The squared Sharpe ratio for $CRF$ is not statistically different from zero at the 5\% significance level. Moreover, the sample squared Sharpe ratio of $MKTS$ is marginally significant at the 5\% level, with other factors such as $DEF$ and $CPTLT$ yielding negative bias-adjusted squared Sharpe ratios. Different from $DEF,$ the $TERM$ factor yields a marginally significant (bias-adjusted) squared Sharpe of 0.014 ($p$-value of 0.077) and seems to be the only (relatively) important driver of the squared Sharpe of the DEFTERM model in Panel~B.

In Panel~B, we consider the sample bias-adjusted squared Sharpe ratios for the BBW, DEFTERM, and HKM models. (The reported $p$-values are based on BKRS.) Strikingly, the four BBW factors jointly produce an even lower bias-adjusted squared Sharpe ratio than $MKTB.$  This finding severely questions the importance of introducing additional bond factors beyond the value-weighted bond market index. Furthermore, the  squared Sharpe ratios of the DEFTERM and HKM factor models are found to be not significantly different from zero at the 5\% level.

Given these surprising findings, we now conduct pairwise model comparison tests based on squared Sharpe ratios following BKRS who focus on a comparison of models' maximum squared Sharpe ratios in an asymptotic analysis under very general distributional assumptions. The differences in the bias-adjusted squared Sharpe ratios (row model minus column model) with their associated \textit{p}-values are presented in Panel~C of Table~2.
The large \textit{p}-values suggest that all of the models perform similarly. The bias-adjusted squared Sharpe ratio of CAPMB (the single-factor model with the $MKTB$ factor) is higher than that of any other factor model considered. For example, the bias-adjusted squared Sharpe ratio of simply holding the value-weighted bond market factor is slightly  higher than that of holding all four of the BBW factors, with a squared Sharpe ratio difference of 0.001 and a $p$-value of 0.216.\footnote{In contrast, based on the original BBW factors, the difference in bias-adjusted squared Sharpe ratios between BBW and CAPMB is economically large and strongly statistically significant (0.077 with a $p$-value of 0.000).} These results are consistent with the findings of insignificant factor alphas using the $MKTB$ factor in Panel~A of Table~2, that is, $DRF$, $CRF,$ and $LRF$ are spanned by the bond market factor. Additional tests that were proposed by BKRS suggest that CAPMB is also not dominated in multiple nonnested model comparison ($p$-value of 0.655).

The statistical results and economic intuition outlined in this section are consistent with each other. CAPMB and the BBW four-factor model perform similarly. This key finding suggests that the factors contained in the BBW model (beyond $MKTB$) are not required from an economic and statistical standpoint. From an investor's perspective, it would be more beneficial to simply hold the bond market portfolio and, as a consequence, avoid the large and persistent transaction costs from attempting to rebalance the remaining BBW portfolios at a monthly interval. From a statistical perspective, the additional factors are redundant across a range of tests, including basic GRS alpha tests for nested models and differences in squared Sharpe ratios for nonnested models.

In a recent working paper,  Bai, Bali, and Wen (2023) acknowledge that $DRF,$ $CRF,$ and $LRF$ are spanned by $MKTB.$ Their newly-proposed four-factor model, which includes short-term reversal ($REV$) in addition to $DRF,$ $CRF,$ and $LRF,$ is shown to perform better than the bond CAPM.
In the Internet Appendix, we demonstrate  that the remarkable performance of $REV$ is essentially due to the pervasive presence of microstructure noise in the WRDS and TRACE data. Once the $REV$ series is purged of market microstructure noise, which is pervasive in WRDS due to transaction-based prices, we cannot reject the null hypotheses that its premium and performance, as measured by the squared Sharpe ratio metric, are indistinguishable from zero. Moreover, the four-factor model with the correctly constructed $REV$ in addition to
 $DRF,$ $CRF,$ and $LRF$ has an associated tangency portfolio squared Sharpe ratio of only 0.010 and is not significantly different from zero ($p$-value of 0.226).

Overall, is it really necessary to utilize ad hoc four-factor models instead of CAPMB in corporate bond pricing? Initial evidence suggests that this may not be the case. In the following analysis, we will dig deeper and report goodness-of-fit measures and risk premia estimates from two-pass CSRs. This additional analysis is motivated by the observation that a model's risk premia do not necessarily coincide with the factor means unless the model is correctly specified, that is, the fundamental beta-pricing restriction holds.

\section{Goodness-of-fit measures and risk premia}\label{sec:fit}

This section reports cross-sectional asset pricing results for the various
 traded-factor models and test portfolio returns described above.
 In Table~3, we report prices of multivariate beta risk and covariance risk ($\gamma$s and $\lambda$s, respectively) and goodness-of-fit measures ($\text{R}^2$s) for the considered beta-pricing models.\footnote{Let $V_f=\text{Cov}(f),$ $\gamma=[\gamma_0,\; \gamma_f']',$ and $\lambda=[\lambda_0,\; \lambda_f']',$ where $\gamma_0 = \lambda_0$ denotes the zero-beta rate in the CSR, and $\gamma_f=V_f\lambda_f$ and $\lambda_f$ represent the $K$-vectors of prices of multivariate beta risk and covariance risk associated with factors $f,$ respectively. It is easy to verify that the pricing errors in the two CSRs are the same (and so are the $\text{R}^2$s). Moreover, since the analysis is in terms of excess returns, we test whether $\gamma_0=\lambda_0=0.$ A preliminary analysis of the rank of the $\beta$ matrix (augmented with a vector of ones to capture the presence of the zero-beta rate in the second-pass) confirms that all models are well-identified and suggests that the factors in each model are individually and jointly not weak. In this context, the quantities of interest (prices of beta/covariance risk and $\text{R}^2$s) are well defined and inference on these parameters can be carried out based on the usual critical values of the tests.}

\begin{center}
\fbox{Table~3 about here}
\end{center}
 We include the OLS and GLS CSR $\text{R}^{2}$s  and their corresponding $p$-values for the test of the null hypothesis that the true $\text{R}^{2}$ is equal to 1, i.e., the model is correctly specified. (See Kan, Robotti, and Shanken, 2013, henceforth KRS, for details.)
 As for the prices of beta/covariance risk, we report the $t$-statistic under correctly specified models that accounts for the errors-in-variables problem ($t$-stat$_{c}$) and the
misspecification-robust $t$-statistic ($t$-stat$_{m}$) proposed by KRS.\footnote{
The $t$-statistic under correctly specified models is the standard
generalized method of moments $t$-statistic under conditional
heteroskedasticity and serial correlation.}

Consistent with the findings of LNS and Kleibergen and Zhan (2015), Panel~A shows that the OLS CSR $\text{R}^2$s are found to be often unrealistically large, ranging from 0.839 to 0.927, and for all of the models we cannot reject the null of exact pricing. However, it is well known that the OLS CSR $\text{R}^2$ has little economic interpretation and can be large even when the fundamental
asset pricing relation is violated. (See Kandel and Stambaugh, 1995.) Nonetheless, notice that CAPMB's $\text{R}^2$ of 0.888 is fairly close to the
value of 0.927 for BBW, and even the most powerful test will encounter difficulties in discriminating between the two.  As expected, for GLS in Panel~B, we now observe a substantial decrease in $\text{R}^2$ values (ranging from 0.002 for CAPM to 0.185 for BBW), with all of the models being now rejected at the 1\% significance level by the GLS CSR $\text{R}^2$ test.
Overall, the amplified performance of BBW (and of the other models) under an OLS weighting scheme is seriously compromised when considering GLS, with strong rejections at any conventional significance level.

To further emphasize the lack of incremental explanatory power of BBW over CAPMB, in Panel~E we report $\text{R}^2$-based tests of pairwise model comparison. (See KRS for details.)
There is no instance of outperformance of BBW over CAPMB as the relatively small differences in $\text{R}^2$ values translate into very large $p$-values of the tests. For OLS as well as GLS, the differences in $\text{R}^2$ values for all models are so small that lack of power is to be expected. Furthermore, CAPMB is not dominated by the other traded-factor models even in multiple nonnested model comparison tests based on CSR $\text{R}^2$s ($p$-values of 0.728 and 0.590 for OLS and GLS, respectively).

Given the tight relation between the GLS CSR $\text{R}^2$ and the mean-variance frontier, we provide visual evidence of our GLS findings by plotting the mean-standard deviation frontier for the 32 test portfolio returns in Fig.~2.

\begin{center}
\fbox{Figure~2 about here}
\end{center}
In Fig.~2, we include the tangency line for the mean-variance efficient portfolio computed using the 32 basis assets in red. The slope of the green line represents the maximum Sharpe ratio from optimally combining the four BBW factors, while the slope of the blue line is the Sharpe ratio of the bond market factor. (We do not bias-adjust the Sharpe ratio estimates in the figure.) The green and blue lines are practically indistinguishable from each other, providing visual evidence that BBW and CAPMB perform about the same in terms of the Sharpe ratio metric. Furthermore, Fig.~2 indicates that both BBW and CAPMB are very far from achieving mean-variance efficiency, thus confirming the strong model rejections based on the GLS CSR $\text{R}^2.$

In Table~3, we further explore the pricing performance of the various factors by focusing on the price of beta and covariance risk.\footnote{As previously noted by Cochrane (2005) and KRS, it is incorrect to focus on the price of
multivariate beta risk (the gammas) if the factors are correlated and the goal is to determine if an underlying factor is incrementally useful in explaining the cross-section of asset returns. (The factor correlations are sizable in our sample.)
In this case, one needs to consider the price of covariance risk (the lambdas) or, equivalently, the price of univariate beta risk.}
In particular, we consider the price of multivariate beta risk in Panels~A and B, while we examine the possible incremental explanatory power of non-market factors in Panels~C and D.
Based on $t$-stat$_{m}$ and a 5\% significance level of the test,
Panels~A and C  (OLS case) show that $DRF,$ $CRF,$ and $LRF$ are not priced, with the other factors also exhibiting limited explanatory power overall.
For GLS in Panels~B and D, only $LRF$ seems to possess nontrivial (incremental) pricing ability for the cross-section of 32 test portfolios sorted on ratings, credit spreads, maturity, and industry classification.

In summary, when considering the four-factor model of BBW, we find no
evidence of incremental pricing, with the marginal exception of $LRF,$ over the sample period that they study.
It is reassuring to see that our CSR findings largely confirm the economic analysis based on squared Sharpe ratios of the previous section.

\section{Nontraded factors and the cross-section of corporate bond returns}\label{sec:nontraded}
In this section, we consider several nontraded-factor models that have shown some success in pricing the cross-section of excess corporate bond returns.
Consistent with the section for traded-factor models, we first construct mimicking portfolios for the nontraded factors and compare their investment performance using the squared Sharpe ratio metric. Thereafter, we confirm prior results related to performance by showing that models with mimicking portfolios do not outperform CAPMB in terms of squared Sharpe ratios. Finally, similar to the case of traded-factor models, we evaluate pricing and model fit by means of OLS and GLS CSRs at a portfolio level.

\subsection{Nontraded-factor models}

Across the various models considered, most of the factors are publicly available from the various authors' websites. However, the aggregate liquidity factors are not publicly available, and we reconstruct them using the intraday TRACE price and volume data. The alternative models are listed below. To conserve space, we only consider the main specifications in the original papers.

\paragraph{Macroeconomic uncertainty risk model (MACRO)} The role of macroeconomic uncertainty in the cross-section of corporate bonds has been studied by Bali, Subrahmanyam, and Wen (2021). Their proposed two-factor model includes the value-weighted corporate bond market factor, $MKTB,$ and the monthly change in the macroeconomic uncertainty index, $UNC$, constructed by Jurado, Ludvigson, and Ng (2015).

\paragraph{Aggregate liquidity risk models (LIQPS and LIQAM)}
Lin, Wang, and Wu (2011) show that aggregate liquidity risk is a priced risk factor in the cross-section of corporate bond returns. We closely follow their methodology and employ the intraday Enhanced TRACE price and volume data to construct the P\'{a}stor and Stambaugh (2003,  $PS$) and Amihud (2002, $AM$) liquidity factors for corporate bonds. The aggregate liquidity risk factor models consist of the Fama-French three stock market factors (market, $MKTS,$ size, $SMB,$ and value, $HML$), the $DEF$ and $TERM$ factors, and either the $PS$ (LIQPS model) or the $AM$ (LIQAM model) liquidity risk proxies.

\paragraph{Aggregate volatility risk models (VOLPS and VOLAM)}
We follow Chung, Wang, and Wu (2019), who examine the pricing of systematic volatility risk in the cross-section of corporate bonds and posit a seven-factor asset pricing model. The model includes the Fama-French three stock market factors, the $DEF$ and $TERM$ factors, a liquidity risk factor ($PS$ or $AM$ from above), and the first difference in the CBOE VIX ($VIX$).
We denote the aggregate volatility risk factor models by VOLPS when controlling for the $PS$ illiquidity risk factor and by VOLAM when controlling for the $AM$ illiquidity risk factor, respectively.

\paragraph{Nontraded intermediary capital risk model (HKMNT)}
We also consider the two-factor model of He, Kelly, and Manela (2017), which includes the $MKTS$ factor and the nontraded intermediary capital risk factor ($CPTL$).

\paragraph{Long-run consumption risk model (LRC)}
Finally, we investigate whether the EJN long-run consumption factors are priced above and beyond bond market risk.\footnote{We thank Nikolai Roussanov for suggesting this exercise to us.}  EJN show that a one-factor model based on long-run consumption growth, LRC, explains the risk premiums on corporate bond portfolios sorted on credit rating, credit spreads, downside
risk, idiosyncratic volatility, long-term reversals, maturity, and sensitivity to the financial
intermediary capital factor of HKM.  We purposely keep the analysis of the LRC model separate from that of the other six nontraded-factor models listed above. The main reason is that in the EJN setting, the data is overlapping and a different set of tools is required for appropriate statistical inference.

\subsection{Summary statistics and model comparison tests with factor mimicking portfolios}
We first construct mimicking portfolios by regressing each nontraded factor on a constant and all of the traded-factor returns  considered previously including the two Fama-French $SMB$ and $HML$ factors. Hence, $R = [MKTB, DRF, CRF, LRF, MKTS, SMB, HML, DEF, TERM, CPTLT]$ represents the set of basis asset returns in the analysis.
The $F$-tests of joint significance of the slope coefficients in
the mimicking-portfolio projections yield $p$-values that are all close to 0, which implies that the mimicking portfolios are properly identified.
We present the means, alphas, squared Sharpe ratios, and standard deviations of the mimicking portfolios in Panel~A of Table~4.

\begin{center}
\fbox{Table~4 about here}
\end{center}
 Only $UNCM,$ the mimicking portfolio for $UNC,$ has a statistically significant mean of $-$0.303\% per month at the 5\% level. (See BKRS for details on how to perform inference on mimicking portfolio risk premia.) Notably, neither $PSM$ nor $AMM$ (the mimicking portfolios for $PS$ and $AM$) yield statistically significant premia, which confirms the findings of Goldberg and Nozawa (2021), who also do not reject the null hypotheses of zero Amihud and P\'{a}stor-Stambaugh factor premia. Similar to Table~2, we also report alpha-based tests for factor mimicking portfolios and investigate whether each of the five mimicking portfolios produces a statistically significant alpha when regressed on $MKTB.$\footnote{Note that as with traded-factor models, testing the equality of squared Sharpe ratios of mimicking portfolios when the two models are nested amounts to evaluating the
hypothesis that the alphas of the mimicking portfolios excluded from the smaller
model are zero when regressed on the mimicking portfolios common to both
models. In this case, however, we can no longer use a basic alpha-based test because we have generated
regressors (the mimicking portfolio weights). Fortunately, one can use the results in Proposition~3 and subsequent discussion of BKRS to obtain the $p$-values associated with the alphas.} All of the alphas are found to be not statistically different from zero, which implies that the mimicking portfolios for the various nontraded factors ($UNCM,$ $CPTLM,$ $PSM,$ $AMM,$ and $VIXM,$ respectively) are spanned by the bond market factor.
Moreover, only the bias-adjusted squared Sharpe ratio of $UNCM$ is significantly different from zero at the 5\% significance level, with three out of four of the other mimicking portfolio bias-adjusted squared Sharpe ratios being negative and not significantly different from zero.\footnote{For mimicking portfolio squared Sharpe ratios, we cannot simply apply the same bias adjustment of the traded-factor case. The reason is that the mimicking portfolios are estimated, and one needs to take into account the estimation uncertainty in the weights of the mimicking portfolios. Therefore, following BKRS, we present jackknife bias-adjusted squared Sharpe ratios.}

In Panel~B, we report the bias-adjusted squared Sharpe ratios for the previously described models with traded factors as well as mimicking portfolios in place of the nontraded factors. Although all of the bias-adjusted squared Sharpe ratios are significantly different from zero at the 10\% significance level, they turn out to be relatively small in magnitude and some of them are even negative, as for VOLPS and VOLAM. Importantly, as shown in Panel~C, CAPMB is never outperformed in pairwise model comparison tests by any of these models with traded factors as well as mimicking portfolios in place of their nontraded counterparts.
 Furthermore, CAPMB is not dominated by the other nontraded-factor models even in multiple nonnested model comparison tests based on mimicking-portfolio squared Sharpe ratios. (For this scenario, the $p$-value of the multiple nonnested model comparison test based on BKRS is 0.751.) These findings are very similar to those for the traded-factor models in Table~2.

 Finally, in Panel~D, we investigate the performance of the one-factor model of EJN. We report results for their conditional long-run risk measure since this is EJN's preferred proxy for the wealthy households' long-run consumption
risk.\footnote{The results for the unconditional long-run risk measure of EJN are qualitatively similar to those based on its conditional counterpart and are available from the authors upon request.} The analysis is based on monthly data with quarterly compounding as in Table~12 of EJN. Specifically, we form a consumption mimicking portfolio by projecting long-run consumption growth on some basis asset returns over our sample period (2004:08-2016:12) and EJN's sample period (1984:03-2019:12). We consider two sets of basis assets: (i) six bond portfolios independently sorted on three maturity bins and two credit rating bins (EJN basis assets) and (ii) the ten  basis assets listed at the beginning of the section (DMR basis assets). We then focus on the mimicking portfolio mean and the alpha from regressing the mimicking portfolio on $MKTB$ (means and alphas are quarterly figures and are multiplied by 100 for ease of exposition). Given the overlapping nature of the data, we employ the nonparametric block bootstrap proposed by EJN to compute the standard errors of the mean and alpha estimates. We report two types of bootstrap standard errors computed with 5,000 replications: (i) the EJN standard errors ($se_{\text{EJN}}$) and (ii) our standard errors ($se_{\text{DMR}}$) that account for the estimation error in the mimicking portfolio
weights.\footnote{EJN directly resample the factor-mimicking portfolio in each bootstrap iteration. Their method neglects the large estimation error in the mimicking portfolio weights and produces bootstrap standard errors that are too small. In contrast, in the standard error computation, we jointly resample the raw data and re-estimate the factor-mimicking portfolio at each bootstrap iteration. This procedure properly accounts for the estimation error in the mimicking portfolio weights and alphas.} When focusing on the shorter sample, we cannot reject the null hypotheses that the mimicking portfolio means and alphas are zero, regardless of the type of standard error and set of basis assets that are considered. Over the longer sample, we confirm the results in EJN when inference is based on their standard errors, but we find that the consumption mimicking portfolio is spanned by the bond market factor when inference is based on our standard error that accounts for the variability in the composition of the mimicking portfolio.
Given this preliminary evidence that the consumption risk factor
may be subsumed by the bond market factor and the inherent challenges in the statistical implementation of two-pass CSRs when the data is overlapping, we drop the LRC model from the subsequent analysis.\footnote{See Gospodinov and Robotti (2021a) for a thorough analysis of two-pass cross-sectional regressions with overlapping data.}

\subsection{Price of beta/covariance risk and CSR $\text{R}^2$s}

Instead of forming mimicking portfolios, we now analyze the performance of the previously described nontraded-factor models by means of OLS and GLS CSRs. Similar to the traded-factor case,  we first subject the various models to rank tests to determine whether they are well identified. Next, we investigate whether the various factors have incremental pricing ability for our cross-section of 32 bond portfolio returns.

 Based on the results of the (approximate) finite-sample $F$-test proposed by Kan and Robotti (2012), we cannot reject the null of reduced rank of the augmented beta matrix for five out of six models.\footnote{To conserve space, the results of the rank tests are only available from the authors upon request.} Only the macro uncertainty model (MACRO) appears to be well-identified. This is a scenario where the OLS risk premium results must be interpreted with caution. In contrast, misspecification-robust GLS inference remains asymptotically valid even when some or all of the factors in an asset pricing specification are weak. (See Gospodinov and Robotti, 2021b, for details.)

Furthermore, we assess whether the factors in the above empirical specification command statistically significant and large risk premia by focusing on the price of beta and covariance risk in Table~5.

\begin{center}
\fbox{Table~5 about here}
\end{center}
While we estimate and test the six full specifications described above, we only report a reduced set of estimate in Panels~A through D of Table~5 to conserve space. Specifically, for each model, we report the parameter estimates for the appropriate market factor ($MKTB$ in MACRO and $MKTS$ in the other five models) and for the nontraded factor of interest.
Regardless of OLS vs. GLS and price of beta vs. price of covariance risk, the results in the table indicate that none of the factors in the various models are priced in the cross--section of bond excess returns.
As for pairwise model comparison tests in Panel~E, the overall conclusions are very similar to those for traded-factor models. The large $p$-values of the tests indicate that CAPMB is never outperformed by any nontraded-factor models in terms of OLS or GLS CSR $\text{R}^2$s.
Moreover, CAPMB is not dominated by factor models with traded and nontraded factors even in multiple nonnested model comparison tests based on CSR $\text{R}^2$s ($p$-values of 0.327 and 0.631 for OLS and GLS, respectively).

For traded- as well as nontraded-factor models, we also perform the analysis based on alternative sets of test portfolios: 25 portfolios sorted on size and ratings, 25 portfolios sorted on credit spreads, 25 portfolios sorted on size and maturity, and 30 Fama-French industry portfolios (the latter two sets of portfolios were also considered by BBW in their cross-sectional analysis). The results from these additional empirical investigations largely confirm our main findings and are available from the authors upon request.

\section{Bond-level analysis}\label{sec:bond_ap}
To investigate, at the bond level, whether the previously-described traded and nontraded factors capture systematic variation in corporate bond returns and lead to sizeable risk premia, we run Fama-MacBeth CSRs using post-ranking factor betas as in Fama and French (1992).
The use of post-ranking factor betas in the CSR is meant to reduce the estimation error in the betas and the attenuation bias in the risk premium estimates, as emphasized by BNS and Goldberg and Nozawa (2021), among others.
 Post-ranking factor betas are estimated at the portfolio level by regressing post-ranking
portfolio excess returns on factor returns and are assigned to each bond in the portfolio. Post-ranking
portfolio excess returns are formed by sorting on pre-ranking factor betas, which are estimated over rolling 36-month windows with a minimum of 24 months required to include the coefficient estimate in the sample.\footnote{This procedure yields a time series of pre-ranking factor betas for each bond in the sample spanning the period 2006:08 to 2016:12 (125 months).}  Consistent with BNS, we form 5 post-ranking equally-weighted portfolios based on a factor's pre-ranking betas in a given model.\footnote{For multifactor models, with the only exception of BBW, we include all factors of a given model in the time-series regressions to obtain pre- and post-ranking factor betas. As for the BBW model, following the original article and BNS, we estimate pre- and post-ranking factor betas in a slightly different way. Pre- and post-ranking betas for the $MKTB$ factor are obtained in isolation based on the single-factor bond CAPM. In contrast, the pre- and post-ranking betas for $DRF,$ $CRF,$ and $LRF$ are obtained from bivariate time-series regressions that
include $MKTB.$} To keep up with the main theme of the paper and determine whether a factor adds to the model's explanatory power at a bond level, we also consider an alternative setting where we replace pre- and post-ranking betas with pre- and post-ranking covariances, while leaving everything else unchanged.\footnote{Because of the way post-ranking portfolio returns are formed, we lose the equivalence between $\gamma_0$ and $\lambda_0,$ and the $\text{R}^2$s of the two CSRs now differ from each other.}

 Table~6 displays the time-series average of the intercept and slope coefficients, and the average adjusted $\text{R}^2$ values. As in BNS, the Fama-MacBeth \textit{t}-statistics are based on a 12-lag NW adjustment.\footnote{As for the portfolio-level analysis for nontraded-factor models in Table~5, in the interest of brevity, we only report a reduced set of estimates that involve the nontraded factor of interest and the appropriate market factor ($MKTB$ in MACRO and $MKTS$ in the other five models).}

\begin{center}
\fbox{Table~6 about here}
\end{center}

\paragraph{Traded-factor models} Panel~A shows that the $MKTB$ and $LRF$ factors in BBW generate significant beta premia of 0.36\% and 0.15\% per month (with \textit{t}-statistics of 1.97 and 2.03, respectively). $LRF$ continues to maintain some nontrivial incremental pricing ability when focusing on the price of covariance risk in Panel~B.  These results confirm the portfolio-level GLS findings, where $LRF$ also plays an important role in explaining the cross-section of bond returns.
There is no evidence of (incremental) pricing for any other traded factor in the analysis.

\paragraph{Nontraded-factor models} We report the Fama-MacBeth risk premia estimates in Panels~C and D of Table~6.
With the marginal exception of the P\'{a}stor-Stambaugh nontraded liquidity factor in Panel~D, we cannot reject the null of zero risk premia for the various nontraded factors in the analysis. This confirms and reinforces the findings of Goldberg and Nozawa (2021), who show that neither $PS$ nor $AM$ are priced at an individual bond level.

For robustness, we also employ the methods of Gagliardini, Ossola, and Scaillet (2016)  to perform bond-level analysis. The main findings from this additional analysis are consistent with those from Fama and MacBeth (1973) regressions. Finally, although the spurious factor problem described earlier should not be a major concern for the Fama-MacBeth regressions given the small time-series sample size in the rolling-window estimation of the betas/covariances, in unreported experiments based on portfolio sorts (which are robust to factor spuriousness) we confirm the main message from Fama-MacBeth CSRs.\footnote{It is well-known that when the time-series sample size is small, the spurious factor problem is not of first order importance. See, for example, recent contributions on the topic by Kleibergen and Zhan (2023) and Kroencke and Thimme (2023).}

The various investigations at a factor-, portfolio-, and bond-level indicate that the proposed bond factors/models are inadequate in explaining the cross-sectional variation in bond returns over the sample period considered by BBW and others. In unreported analyses, we also extended the sample period back to January 1986 in the attempt of increasing the power of our tests. The results are remarkably similar to those for the shorter sample considered in the paper. In the Internet Appendix, we consider two alternatives to TRACE, that is, the Wharton Research Data Services (WRDS) and the Intercontinental Exchange (ICE) bond return databases. When applying our battery of tests to these newly constructed factors and returns, the evidence of (incremental) pricing ability for the various factors becomes even weaker, although traded liquidity continues to perform marginally better than the other factors.
Overall, robust evidence for common factor pricing in corporate bonds remains elusive.

\section{Conclusion}\label{sec:conclusion}
A prominent trend in recent empirical asset pricing research has been the persistent search for risk factors that demonstrate robust pricing performance.
Despite this focus, assessing factor models on test assets, even within a single asset class such as equities or bonds, remains difficult due to various issues such as model uncertainty, poor model identification, small time series sample sizes relative to the number of test assets, and more. These issues are even more fraught at the firm or individual asset level. This is particularly true in the case of corporate bonds, where short sample periods and unreliable publicly available factors exacerbate these issues. Within the context of corporate bonds, the relentless search for factors has been applied verbatim from the equity literature.

In this article, we explore the limitations of evaluating factor models on corporate bonds, specifically within the context of the Bai, Bali, and Wen (2019) four-factor model and other models with traded and nontraded factors. Our analysis highlights the challenges associated with assessing the economic and statistical significance of the proposed risk factors, and we offer recommendations to create a reliable framework for this evaluation. Overall we find that it is difficult for newly proposed specifications to outperform the simple bond CAPM, economically and statistically.
Our results are robust to a variety of checks, including portfolio- vs. bond-level analysis, excess vs. duration-adjusted returns, and shorter vs. longer sample periods.
Further work on frequency as
a dimension of risk, along the lines of Bandi, Chaudhuri, Lo, and Tamoni (2021) and  Neuhierl and Varneskov (2021) for equities, may
 provide valuable groundwork for a better understanding of the cross-sectional determinants of expected corporate bond returns. Finally,
 given the nontrivial transaction costs in the over-the-counter trading of corporate bonds, it would be valuable to formally compare the performance of alternative pricing models for bonds based on economically meaningful metrics that take into account transaction costs along the lines of Detzel, Novy-Marx, and Velikov (2023) for equities.

\renewcommand\thesection{\Alph{section}}
\setcounter{section}{0}

\clearpage
\section*{Appendix}
The Enhanced TRACE (TRACE) bond database  is the primary database used in all of the papers that we revisit. TRACE provides intraday bond clean prices, trading volumes, and buy-and-sell indicators.
We apply the standard bond filtering procedure used in the original BBW paper. The filters primarily use bond characteristics from the FISD database and variables contained in the intraday TRACE bond database.

\paragraph{TRACE bond filters}
We apply the following filters in cleaning the intraday TRACE data.

\begin{enumerate}
    \item Keep all trades that have less than two days to settlement, \code{days\_to\_sttl\_ct == `002'},\\
    \code{days\_to\_sttl\_ct == `001'}, \code{days\_to\_sttl\_ct == `000'} or \code{days\_to\_sttl\_ct == `None'}.

    \item Remove trade records with the `when-issued' indicator, \code{wis\_fl != `Y'}.

    \item Remove trade records with the `locked-in' indicator, \code{lckd\_in\_ind != `Y'}.

    \item Keep trade records which do not have special conditions, \code{sale\_cndtn\_cd == `None'} or \code{sale\_cndtn\_cd == `@'}.

    \item Keep trades that register a daily par volume equal or greater than \$10,000, \code{entrd\_vol\_qt >= 10000}.

     \item Keep trades that include bond prices less than \$1,000 and greater than \$5, \code{(rptd\_pr > 5) \& (rptd\_pr < 1000)}. The removal of this filter has no material impact on the results.
\end{enumerate}

Thereafter, we clean the bond trades for reversals, corrections, and cancellations in the standard manner as prescribed by Dick-Nielsen (2014). The end-of-day bond clean price is the volume-weighted price of all eligible trades within each day $d$ of month $t$.

\clearpage
\paragraph{FISD bond filters}
\begin{enumerate}
    \item Only keep bonds that are issued by firms domiciled in the United States of America, \code{COUNTRY\_DOMICILE == "USA"}.

    \item Remove bonds that are private placements, \code{PRIVATE\_PLACEMENT == "N"}.

    \item Only keep bonds that are traded in U.S. Dollars, \code{FOREIGN\_CURRENCY == "N"}.

    \item Bonds that trade under the 144A Rule are discarded, \code{RULE\_144A == "N"}.

    \item Remove all asset-backed bonds, \code{ASSET\_BACKED == "N"}.

    \item Remove convertible bonds, \code{CONVERTIBLE == "N"}.

    \item Only keep bonds with a fixed or zero coupon payment structure, i.e., remove bonds with a floating (variable) coupon, \code{COUPON\_TYPE != "V"}.

    \item Remove bonds that are equity linked, agency-backed, U.S. Government, and mortgage-backed, based on their \code{BOND\_TYPE}.

       \item Remove bonds that have a ``non-standard'' interest payment structure or bonds not caught by the variable coupon filter (\code{COUPON\_TYPE}). This affects a tiny fraction of bonds ($\sim 0.10\%$ or 142 bonds) of the FISD data file. We remove bonds that have an  \code{INTEREST\_FREQUENCY} equal to $-$1 (N/A), 13 (Variable Coupon), 14 (Bi-Monthly), and 15 and 16 (undocumented by FISD). Additional information on  \code{INTEREST\_FREQUENCY} is available on Page 60 of 67 of the FISD Data Dictionary 2012 document.

       \item Remove a small fraction of bonds that do not have the required (and crucial information) to compute accrued interest. Bonds that do not have a valid \code{DATED\_DATE} are removed (3,051 bonds). The \code{DATED\_DATE} variable is the date from which bond interest accrues. Bonds without a valid \code{INTEREST\_FREQUENCY}, \code{DAY\_COUNT\_BASIS}, \code{OFFERING\_DATE}, \code{COUPON\_TYPE}, and \code{COUPON} are also removed (425 bonds in total).
\end{enumerate}

For bonds with missing amount outstanding information in the file, we set the amount outstanding equal to the face value at issuance.

\paragraph{Sample coverage}
In Panel A of Table~A1, we present the total number of bond-month observations, bonds, firms, and the average number of bonds and firms in any given month $t$ throughout the sample. A bond or firm observation is valid if it has non-missing data for the bond return, bond rating, and bond amount outstanding.

\begin{center}
		\fbox{Table~A1 about here}
\end{center}
By strictly following the BBW sampling procedure for the TRACE data, we are only able to produce 69\% of their bond-month observations, 80\% of their bonds, and 93\% of their firms.
We further investigate these findings in Panel~B. We expand the number of days that a valid return can be computed with based on the TRACE database. We use a number of days equal to 1 (end of the month), 3, 5 (as in Panel~A), 10, and $>10$ (i.e., any day of the month). For $n>10$ (a return is computed if the bond traded on \textit{any} day of the month),  we obtain numbers that are very similar to what BBW report.

\paragraph{Summary statistics}
In Panel~A of Table~A2, we report the time-series average of the cross-sectional mean, median, standard deviation, and percentiles for various bond characteristics that are used in data construction. In Panel~B, we report their average correlations. We do not winsorize any of the variables, except for $ILLIQ,$ which has a few extreme outliers. This variable is winsorized at the 1\% level (0.50\% in each tail).

\begin{center}
		\fbox{Table~A2 about here}
\end{center}

\clearpage

\clearpage
\begingroup
\fontsize{11pt}{19pt}\selectfont
\begin{table}[h!]
\textbf{Table~1} \newline  \footnotesize{Bai, Bali, and Wen (2019) four-factor comparison. \\[0.04in] \hspace*{0.1in}
Panels~A and B report factor means (Mean), standard deviations (SD), and percentiles for the original and replicated factors, respectively.
The factors include the bond market factor (\textit{MKTB}), the downside risk factor (\textit{DRF}), the credit risk factor (\textit{CRF}), and the liquidity risk factor (\textit{LRF}). Panel~C.1 reports the {pairwise} correlations between the original factors (which contain lead-lag errors) and the replicated factors, and Panel~C.2 reports the  {pairwise} correlations between the corrected original factors and the replicated factors. Panels~D.1 and D.2 report the pairwise correlations for the original factors and the replicated factors, respectively. Panels~A--D are based on the sample period 2004:08 to 2016:12 (149 months).}
\vspace{0.1in} \par
\label{tab:Table_001}
\begin{tabular*}{\textwidth}{@{\extracolsep{\fill}}lccccccccc}
\toprule
\multicolumn{10}{c}{Panel A: Original factors} \\ \midrule
 & Mean & SD & Median & Min & 5th & 25th & 75th & 95th & Max \\ \midrule
\textit{MKTB} & 0.333 & 1.381 & 0.410 & $-6.365$ & $-1.298$ & $-0.467$ & 0.987 & 2.299 & 7.568 \\
\textit{DRF}  & 0.694 & 2.381 & 0.595 & $-7.430$ & $-2.451$ & $-0.553$ & 1.722 & 4.462 & 12.789 \\
\textit{CRF}  & 0.431 & 1.876 & 0.327 & $-8.839$ & $-2.354$ & $-0.429$ & 1.357 & 3.089 & 8.194 \\
\textit{LRF}  & 0.491 & 1.418 & 0.321 & $-2.629$ & $-0.936$ & $-0.243$ & 0.938 & 2.230 & 11.660 \\
              &       &       &       &          &          &          &       &       &  \\
\multicolumn{10}{c}{Panel B: Replicated  factors} \\ \midrule
\multicolumn{1}{c}{} & Mean & SD & Median & Min & 5th & 25th & 75th & 95th & Max \\ \midrule
\textit{MKTB} & 0.469 & 1.892 & 0.495 & $-9.292$ & $-1.856$ & $-0.460$ & 1.310 & 2.832 & 10.809 \\
\textit{DRF}  & 0.673 & 3.355 & 0.633 & $-15.895$ & $-3.436$ & $-0.618$ & 1.702 & 4.724 & 16.768 \\
\textit{CRF}  & 0.508 & 3.411 & 0.531 & $-21.908$ & $-3.402$ & $-0.889$ & 2.174 & 4.950 & 13.233 \\
\textit{LRF}  & 0.361 & 1.470 & 0.269 & $-5.078$ & $-1.447$ & $-0.295$ & 0.759 & 2.461 & 8.719 \\
 &  &  &  &  &  &  &  &  &  \\
\end{tabular*}

\begin{tabular*}{\textwidth}{@{\extracolsep{\fill}}lccccccccc}
\multicolumn{10}{c}{Panel C: Pairwise correlations across factors} \\ \midrule
\multicolumn{5}{c}{C.1: Original-replicated} & \multicolumn{1}{l}{} & \multicolumn{4}{c}{C.2: Original (corrected)-replicated} \\  \cmidrule(lr){2-5}  \cmidrule(lr){6-10}
& \textit{MKTB} & \textit{DRF} & \textit{CRF} & \textit{LRF} &  & \textit{MKTB} & \textit{DRF} & \textit{CRF} & \textit{LRF} \\\cmidrule(lr){2-5}  \cmidrule(lr){6-10}
\textit{MKTB} & 0.939 &  &  &  &  & 0.939 &  &  &  \\
\textit{DRF} &  & 0.264 &  &  &  &  & 0.931 &  &  \\
\textit{CRF} &  &  & 0.445 &  &  &  &  & 0.948 &  \\
\textit{LRF} &  &  &  & 0.829 &  &  &  &  & 0.880 \\
 &  &  &  &  &  &  &  &  &  \\
\end{tabular*}

\begin{tabular*}{\textwidth}{@{\extracolsep{\fill}}lccccccccc}
\multicolumn{10}{c}{Panel D:   Pairwise correlations across factors} \\ \midrule
\multicolumn{5}{c}{D.1: Original} &  & \multicolumn{4}{c}{D.2: Replicated} \\ \cmidrule(lr){2-5}  \cmidrule(lr){6-10}
& \textit{MKTB} & \textit{DRF} & \textit{CRF} & \textit{LRF} &  & \textit{MKTB} & \textit{DRF} & \textit{CRF} & \textit{LRF} \\\cmidrule(lr){2-5}  \cmidrule(lr){6-10}
\textit{MKTB} & 1 & 0.284 & 0.455 & 0.470 &  & 1 & 0.785 & 0.455 & 0.618 \\
\textit{DRF} &  & 1 & 0.424 & 0.319 &  &  & 1 & 0.381 & 0.803 \\
\textit{CRF} &  &  & 1 & 0.352 &  &  &  & 1 & 0.411 \\
\textit{LRF} &  &  &  & 1 &  &  &  &  & 1\\  \bottomrule
\end{tabular*}
\end{table}
\endgroup

\clearpage
\begingroup
\fontsize{11pt}{19pt}\selectfont
\begin{table}[p]
\textbf{Table~2} \newline  \footnotesize{Summary statistics and performance tests for traded-factor models. \\[0.04in] \hspace*{0.1in}
Panel~A reports means (Mean), the single-factor bond market alphas (Alpha), the  bias-adjusted factor squared Sharpe ratios (Sh$^2$), and the standard deviations (SD) for the bond market factor (\textit{MKTB}), the downside risk factor (\textit{DRF}), the credit risk factor (\textit{CRF}), the liquidity risk factor (\textit{LRF}), the default risk factor (\textit{DEF}), the term structure factor (\textit{TERM}), the stock market factor (\textit{MKTS}), and the traded version of the He, Kelly, and Manela (2017) intermediary capital risk factor (\textit{CPTLT}). Panel~B reports the bias-adjusted model squared Sharpe ratios for the four-factor model of Bai, Bali, and Wen (2019, BBW), the two-factor model with default and term structure factors of Fama and French (1993, DEFTERM), and the two-factor model of He, Kelly, and Manela (2017, HKM).
In Panel~C, we report pairwise nested and nonnested model comparison tests based on bias-adjusted squared Sharpe ratios (the differences are computed between the squared Sharpe ratios in row $i$ and column $j$). The additional models in Panel~C include the CAPM with the corporate bond market factor (CAPMB), the equity CAPM with the stock market factor (CAPM), and
the single-factor model of HKM (HKMSF).
Panels~A--C are based on the sample period 2004:08 to 2016:12 (149 months). $p$-values are in square brackets.}
\vspace{0.1in} \par
\label{tab:Table_002}
\begin{tabular*}{\textwidth}{@{\extracolsep{\fill}}lcccccccc}
\toprule
    \multicolumn{9}{c}{Panel~A: Factor statistics and squared Sharpe ratios} \\  \midrule
& \textit{MKTB} & \textit{DRF} & \textit{CRF} & \textit{LRF} & \textit{DEF} & \textit{TERM} & \textit{MKTS} & \textit{CPTLT} \\ \midrule
Mean  & 0.469   & 0.673   & 0.508   & 0.361   & 0.020        & 0.478   & 0.675   & 0.502 \\
      & [0.009] & [0.023] & [0.163] & [0.015] & [0.907]      & [0.064] & [0.073] & [0.463] \\
Alpha & -       & 0.020   & 0.123   & 0.135   & $-$0.313     & 0.307   & 0.156   & $-$0.149 \\
      & -       & [0.932] & [0.681] & [0.130] & [0.038]      & [0.261] & [0.622] & [0.789] \\
Sh$^2$ & 0.054   & 0.033   & 0.015   & 0.052   & $-$0.007 & 0.014   & 0.019   & $-$0.002 \\
       & [0.002] & [0.014] & [0.069] & [0.003] & [0.911]  & [0.077] & [0.048] & [0.381] \\
SD     & 1.898   & 3.366   & 3.422   & 1.475   & 2.172    & 3.308   & 4.176   & 7.024 \\
 &  &  &  &  &  &  &  & \\
\end{tabular*}

\begin{tabular*}{\textwidth}{@{\extracolsep{\fill}}lccc}
   \multicolumn{4}{c}{Panel~B: Model squared Sharpe ratios} \\  \midrule
 & BBW & DEFTERM & HKM \\  \midrule
Sh$^2$ & 0.053 & 0.014 & 0.023 \\
 & {[}0.015{]} & {[}0.126{]} & {[}0.061{]} \\
 & & & \\
\end{tabular*}

\begin{tabular*}{\textwidth}{@{\extracolsep{\fill}}lccccc}
  \multicolumn{6}{c}{Panel C: Differences  in model squared Sharpe ratios} \\
\toprule
        & CAPM & HKMSF & HKM & DEFTERM & BBW \\ \midrule
CAPMB & 0.035 & 0.055 & 0.031 & 0.040 & 0.001 \\
      & [0.358] & [0.229] & [0.417] & [0.363] & [0.216] \\
CAPM &  & 0.021 & $-$0.004 & 0.005 & $-$0.034 \\
      & & [0.067] & [0.198] & [0.892] & [0.446] \\
HKMSF &  &  & $-$0.025 & $-$0.015 & $-$0.055 \\
      & &   &  [0.032] & [0.120] & [0.293] \\
HKM &  &  &  & 0.009 & $-$0.030 \\
      & &   &  & [0.808] & [0.512]  \\
DEFTERM &  &  &  &  & $-$0.039\\
      & &   &  & & [0.445] \\ \bottomrule
\end{tabular*}
\end{table}
\endgroup

\newgeometry{left=2.00cm, right=2.00cm, top=2cm, bottom=2.75cm}
\begin{adjustbox}{max width= 1\linewidth,center}
\begin{threeparttable}
\textbf{Table~3} \newline  \footnotesize{Risk premia and CSR R$^2$s for traded-factor models. \\ [0.04in]
\hspace*{0.1in} The table presents the estimation results of six beta-pricing models with traded factors. The models include the CAPM with the corporate bond market factor (CAPMB), the equity CAPM with the stock market factor (CAPM), the four-factor model of Bai, Bali, and Wen (2019, BBW), the two-factor model of He, Kelly, and Manela (2017, HKM), the single-factor model of HKM (HKMSF), and the two-factor model with default and term structure factors of Fama and French (1993, DEFTERM). The models are estimated using monthly excess returns on the 32 combination (combo) portfolios. The sample period is 2004:08 to 2016:12 (149 months).  We report parameter estimates $\hat{\gamma}$ (multiplied by 100) and $\hat{\lambda}$ (with $\hat{\lambda}_0$ multiplied by 100), the  $t$-statistics under correctly specified models  (\textit{t}-$\text{stat}_c$), and the model misspecification-robust \textit{t}-statistics (\textit{t}-$\text{stat}_m$) from Kan, Robotti, and Shanken (2013). In addition, we report the OLS and GLS CSR R$^2$ for each model and the difference between the CSR R$^2$s of the models in row \textit{i} and column \textit{j}. $t$-statistics are in round brackets and $p$-values are in square brackets.}
\vspace{0.1in}
\label{tab:Table_003}
\footnotesize
\setlength\tabcolsep{0.15mm}
\begin{tabular}{lcccccccccccccccccccccc}
\toprule
\multicolumn{23}{c}{Panel A: Price of beta risk (OLS)} \\ \midrule
 & \multicolumn{2}{c}{CAPMB} &  & \multicolumn{5}{c}{BBW} &  & \multicolumn{3}{c}{DEFTERM} &  & \multicolumn{2}{c}{CAPM} &  & \multicolumn{2}{c}{HKMSF} &  & \multicolumn{3}{c}{HKM} \\ \midrule
 & $\hat{\gamma}_{0}$ & $\hat{\gamma}_{MKTB}$ &  & $\hat{\gamma}_{0}$ & $\hat{\gamma}_{MKTB}$ & $\hat{\gamma}_{DRF}$ & $\hat{\gamma}_{CRF}$ & $\hat{\gamma}_{LRF}$ &  & $\hat{\gamma}_{0}$ & $\hat{\gamma}_{DEF}$ & $\hat{\gamma}_{TERM}$ &  & $\hat{\gamma}_{0}$ & $\hat{\gamma}_{MKTS}$ &  & $\hat{\gamma}_{0}$ & $\hat{\gamma}_{CPTLT}$ &  & $\hat{\gamma}_{0}$ & $\hat{\gamma}_{MKTS}$ & $\hat{\gamma}_{CPTLT}$ \\ \midrule
Estimate & 0.01 & 0.48 &  & 0.13 & 0.35 & 0.38 & 0.55 & $-0.02$ &  & 0.18 & 0.57 & $-0.36$ &  & 0.25 & 0.98 &  & 0.30 & 1.71 &  & 0.17 & 1.23 & 0.56 \\
\textit{t}-$\text{stat}_c$ & (0.05) & (1.83) &  & (1.54) & (1.74) & (1.22) & (1.51) & ($-0.09$) &  & (2.82) & (1.57) & ($-0.75$) &  & (1.93) & (1.56) &  & (2.32) & (1.50) &  & (1.42) & (1.75) & (0.44) \\
\textit{t}-$\text{stat}_m$ & (0.05) & (1.80) &  & (1.38) & (1.60) & (1.16) & (1.48) & ($-0.06$) &  & (2.84) & (1.54) & ($-0.80$) &  & (1.94) & (1.56) &  & (2.32) & (1.49) &  & (1.34) & (1.78) & (0.37) \\
R$^2$ & 0.888   & & & 0.927   & & & & & & 0.839   & & & & 0.876   & &  &  0.851   & & &   0.896   & &   \\
      & [0.591] & & & [0.444] & & & & & & [0.307] & & & & [0.500] & &  &  [0.397] & & &   [0.600] & &   \\
 &  &  &  &  &  &  &  &  &  &  &  &  &  &  &  &  &  &  &  &  &  &  \\
 \multicolumn{23}{c}{Panel B: Price of beta risk (GLS)} \\ \midrule
 & \multicolumn{2}{c}{CAPMB} &  & \multicolumn{5}{c}{BBW} &  & \multicolumn{3}{c}{DEFTERM} &  & \multicolumn{2}{c}{CAPM} &  & \multicolumn{2}{c}{HKMSF} &  & \multicolumn{3}{c}{HKM} \\ \midrule
 & $\hat{\gamma}_{0}$ & $\hat{\gamma}_{MKTB}$ &  & $\hat{\gamma}_{0}$ & $\hat{\gamma}_{MKTB}$ & $\hat{\gamma}_{DRF}$ & $\hat{\gamma}_{CRF}$ & $\hat{\gamma}_{LRF}$ &  & $\hat{\gamma}_{0}$ & $\hat{\gamma}_{DEF}$ & $\hat{\gamma}_{TERM}$ &  & $\hat{\gamma}_{0}$ & $\hat{\gamma}_{MKTS}$ &  & $\hat{\gamma}_{0}$ & $\hat{\gamma}_{CPTLT}$ &  & $\hat{\gamma}_{0}$ & $\hat{\gamma}_{MKTS}$ & $\hat{\gamma}_{CPTLT}$ \\ \midrule
Estimate & 0.02 & 0.46 &  & 0.02 & 0.45 & 0.47 & 0.55 & 0.41 &  & 0.02 & 0.15 & $0.22$ &  & 0.03 & 0.15 &  & 0.03 & 0.39 &  & 0.03 & 0.10 & 0.45 \\
\textit{t}-$\text{stat}_c$ & (0.64) & (2.46) &  & (0.67) & (2.43) & (1.57) & (1.52) & (2.41) &  & (0.96) & (0.73) & ($0.78$) &  & (1.09) & (0.31) &  & (1.17) & (0.44) &  & (1.19) & (0.21) & (0.49) \\
\textit{t}-$\text{stat}_m$ & (0.60) & (2.49) &  & (0.69) & (2.43) & (1.54) & (1.52) & (2.37) &  & (0.92) & (0.70) & ($0.76$) &  & (1.04) & (0.29) &  & (1.07) & (0.41) &  & (1.12) & (0.20) & (0.46) \\
 R$^2$ & 0.097   & & & 0.185   & & & & & & 0.026   & & & & 0.002   & &  &  0.004   & & &   0.006   & &   \\
      & [0.002] & & & [0.003] & & & & & & [0.000] & & & & [0.000] & &  &  [0.000] & & &   [0.000] & &   \\
 &  &  &  &  &  &  &  &  &  &  &  &  &  &  &  &  &  &  &  &  &  &  \\
 \multicolumn{23}{c}{Panel C: Price of covariance risk (OLS)} \\ \midrule
 & \multicolumn{2}{c}{CAPMB} &  & \multicolumn{5}{c}{BBW} &  & \multicolumn{3}{c}{DEFTERM} &  & \multicolumn{2}{c}{CAPM} &  & \multicolumn{2}{c}{HKMSF} &  & \multicolumn{3}{c}{HKM} \\ \midrule
 & $\hat{\lambda}_{0}$ & $\hat{\lambda}_{MKTB}$ &  & $\hat{\lambda}_{0}$ & $\hat{\lambda}_{MKTB}$ & $\hat{\lambda}_{DRF}$ & $\hat{\lambda}_{CRF}$ & $\hat{\lambda}_{LRF}$ &  & $\hat{\lambda}_{0}$ & $\hat{\lambda}_{DEF}$ & $\hat{\lambda}_{TERM}$ &  & $\hat{\lambda}_{0}$ & $\hat{\lambda}_{MKTS}$ &  & $\hat{\lambda}_{0}$ & $\hat{\lambda}_{CPTLT}$ &  & $\hat{\lambda}_{0}$ & $\hat{\lambda}_{MKTS}$ & $\hat{\lambda}_{CPTLT}$ \\ \midrule
Estimate & 0.01 & 13.53 &  & 0.13 & 9.83 & $$5.43 & 4.31 & $-23.02$ &  & 0.18 & 12.48 & 0.34 &  & 0.25 & 5.64 &  & 0.30 & 3.50 &  & 0.17 & 16.65 & $-$6.96 \\
\textit{t}-$\text{stat}_c$ & (0.05) & (1.41) &  & (1.54) & (0.70) & (0.41) & (1.07) & ($-0.82$) &  & (2.82) & (1.61) & (0.10) &  & (1.93) & (1.35) &  & (2.32) & (1.43) &  & (1.42) & (1.47) & ($-$1.04) \\
\textit{t}-$\text{stat}_m$ & (0.05) & (1.39) &  & (1.38) & (0.62) & ($$0.30) & (1.01) & ($-0.52$) &  & (2.84) & (1.49) & (0.10) &  & (1.94) & (1.35) &  & (2.32) & (1.42) &  & (1.34) & (1.23) & ($-$0.81) \\
 &  &  &  &  &  &  &  &  &  &  &  &  &  &  &  &  &  &  &  &  &  &  \\
\multicolumn{23}{c}{Panel D: Price of covariance risk (GLS)} \\ \midrule
 & \multicolumn{2}{c}{CAPMB} &  & \multicolumn{5}{c}{BBW} &  & \multicolumn{3}{c}{DEFTERM} &  & \multicolumn{2}{c}{CAPM} &  & \multicolumn{2}{c}{HKMSF} &  & \multicolumn{3}{c}{HKM} \\ \midrule
 & $\hat{\lambda}_{0}$ & $\hat{\lambda}_{MKTB}$ &  & $\hat{\lambda}_{0}$ & $\hat{\lambda}_{MKTB}$ & $\hat{\lambda}_{DRF}$ & $\hat{\lambda}_{CRF}$ & $\hat{\lambda}_{LRF}$ &  & $\hat{\lambda}_{0}$ & $\hat{\lambda}_{DEF}$ & $\hat{\lambda}_{TERM}$ &  & $\hat{\lambda}_{0}$ & $\hat{\lambda}_{MKTS}$ &  & $\hat{\lambda}_{0}$ & $\hat{\lambda}_{CPTLT}$ &  & $\hat{\lambda}_{0}$ & $\hat{\lambda}_{MKTS}$ & $\hat{\lambda}_{CPTLT}$ \\ \midrule
Estimate & 0.02 & 12.82 &  & 0.02 & 18.63 & $-$15.59 & 0.08 & 32.68 &  & 0.02 & 5.60 & $$3.69 &  & 0.03 & 0.88 &  & 0.03 & 0.80 &  & 0.03 & $-2.08$ & 1.94 \\
\textit{t}-$\text{stat}_c$ & (0.64) & (1.81) &  & (0.67) & (1.45) & ($-$1.67) & (0.02) & (2.84) &  & (0.96) & (1.17) & ($$1.30) &  & (1.09) & (0.30) &  & (1.17) & (0.43) &  & (1.19) & ($-0.37$) & (0.54) \\
\textit{t}-$\text{stat}_m$ & (0.60) & (1.82) &  & (0.69) & (1.40) & ($-$1.69) & (0.02) & (2.85) &  & (0.92) & (1.12) & ($$1.27) &  & (1.04) & (0.28) &  & (1.07) & (0.41) &  & (1.12) & ($-0.31$) & (0.46) \\
 &  &  &  &  &  &  &  &  &  &  &  &  &  &  &  &  &  &  &  &  &  &  \\
\end{tabular}

\setlength\tabcolsep{1.25mm}
\begin{tabular}{lcccccccccccccccccccccc}
\multicolumn{23}{c}{Panel E: Differences in CSR R$^2$s} \\ \midrule
\multicolumn{5}{c}{OLS} & \multicolumn{13}{l}{} & \multicolumn{5}{c}{GLS} \\ \midrule
 & BBW & DEFTERM & CAPM & HKMSF & HKM & \multicolumn{11}{l}{} & & BBW & DEFTERM & CAPM & HKMSF & HKM \\ \midrule
CAPMB   & $-0.038$ & 0.049    & $0.012$  & 0.037    & $-0.008$    & & & & & & & & & & & & &  $-0.088$ & 0.071 & 0.095 & 0.093 & 0.091\\
        & [0.816] & [0.471]  & [0.923]  & [0.803]  & [0.929]  & & & & & & & & & & & & &   [0.265] & [0.149] & [0.112] & [0.112] & [0.141]\\
BBW     &          & 0.088    & 0.050    & 0.076    & 0.031    & & & & & & & & & & & & &          & 0.159 & 0.183 & 0.181 & 0.179 \\
        &          & [0.301]  & [0.577]  & [0.516]  & [0.527]  & & & & & & & & & & & & &          & [0.112] & [0.081] & [0.079] & [0.087]\\
DEFTERM &          &         & $-0.037$ & $-0.012$ & $-0.057$ & & & & & & & & & & & & &                  &   & 0.024 & 0.022 & 0.020 \\
        &          &         & [0.791]  & [0.940]  & [0.618]  & & & & & & & & & & & & &                  & & [0.352] & [0.449] & [0.541] \\
CAPM    &          &         &         & 0.025 & $-0.020$ & & & & & & & & & & & & &                      &          &          & $-0.002$ & $-0.004$ \\
        &          &         &         & [0.375]  & [0.415]  & & & & & & & & & & & & &                      &          &          & [0.778] & [0.645] \\
HKMSF   &          &         &         &         & $-0.045$ & & & & & & & & & & & & &                      &          &          &         &$-0.002$\\
        &          &         &         &         & [0.210]  & & & & & & & & & & & & &                      &          &          &
& [0.754] \\ \bottomrule
\end{tabular}
\end{threeparttable}
\end{adjustbox}
\restoregeometry

\clearpage
\begingroup
\fontsize{11pt}{19pt}\selectfont
\begin{table}[h!]
\textbf{Table~4} \newline  \footnotesize{Summary statistics and performance tests for models with mimicking portfolios. \\[0.04in] \hspace*{0.1in}
Panel~A reports the means (Mean), the single-factor bond market alphas (Alpha), the  bias-adjusted factor squared Sharpe ratios (Sh$^2$), and the standard deviations (SD) for the mimicking portfolios on the bond market factor (\textit{MKTB}), the Jurado, Ludvigson, and Ng (2015) uncertainty factor (\textit{UNCM}), the He, Kelly, and Manela (2017) nontraded financial intermediary factor  (\textit{CPTLM}), the P\'{a}stor and Stambaugh (2003) liquidity factor (\textit{PSM}), the Amihud (2002)  liquidity factor (\textit{AMM}), and the volatility factor (\textit{VIXM}). Panel~B reports the bias-adjusted model squared Sharpe ratios for the MACRO and HKMNT two-factor models of Bali, Subrahmanyam, and Wen (2021) and He, Kelly, and Manela (2017), the LIQPS and LIQAM six-factor models of Lin, Wang, and Wu (2011), and the VOLPS and VOLAM seven-factor models of Chung, Wang, and Wu (2019). Panel~C reports the difference between the (bias-adjusted)
squared Sharpe ratio of CAPMB and those of the given models with the associated $p$-value for the test of equality (0 difference).  Panels~A--C are based on the sample period 2004:08 to 2016:12 (149 months). In Panel~D, we report the means and the single-factor bond market alphas (multiplied by 100) for the mimicking portfolios on the conditional long-run consumption factor of Elkamhi, Jo, and Nozawa (2023), based on our sample (2004:08-2016:12) vs. the EJN sample (1984:03-2019:12) and our set of basis assets (DMR basis assets) vs. the EJN set of basis assets (EJN basis assets). We report two types of bootstrapped standard errors
computed with 5,000 replications: (i) the EJN standard errors ($se_{\text{EJN}}$) and (ii) our standard errors ($se_{\text{DMR}}$) that account for the estimation error in the mimicking portfolio weights.
In the table, the $p$-values are in square brackets and the standard errors are in round brackets.}
\vspace{0.1in} \par
\centering
\label{tab:Nontraded_mimicking}
\begin{tabular}{lcccccc}
\toprule
    \multicolumn{7}{c}{Panel   A: Mimicking portfolio statistics and squared Sharpe ratios} \\  \midrule
       & \textit{MKTB} & \textit{UNCM} & \textit{CPTLM} & \textit{PSM} & \textit{AMM} & \textit{VIXM}  \\ \midrule
Mean   & 0.469         & $-0.303$     & 0.369        & $-0.094$   & 0.147      & $-0.497$  \\
       & [0.009]       &[0.042]       & [0.579]      & [0.592]    & [0.411]    & [0.117]   \\
Alpha  & -             & $-0.084$     & $-0.164$     & $-0.268$   & $-0.052$   & $0.105$  \\
       & -             &[0.411]       & [0.758]      & [0.177]    & [0.761]    & [0.692]   \\
Sh$^2$ & 0.054         &0.064         & $-0.007$     & $-0.057$   & $-0.013$   & $0.023$  \\
       & [0.002]       &[0.004]       & [0.478]      & [0.505]    & [0.372]    & [0.077]   \\
SD     & 1.898         &1.284         & 6.367        & 1.718      & 2.021      & 3.438     \\
       &               &              &              &            &            &           \\
\end{tabular}

\begin{tabular}{lcccccc}
   \multicolumn{7}{c}{Panel B: Model squared Sharpe ratios} \\  \midrule
       & MACRO     & HKMNT         & LIQPS         & LIQAM        & VOLPS         & VOLAM       \\  \midrule
Sh$^2$ & 0.047     & 0.028         & 0.057         & 0.032        & $-0.048$      & $-0.056$    \\
       & [0.006]   & [0.048]       & [0.052]       & [0.053]      & [0.084]       &  [0.065]    \\
       &           &               &               &              &               &             \\
\end{tabular}

\setlength\tabcolsep{1.50mm}
\begin{tabular}{lcccccc}
\multicolumn{7}{c}{{Panel~C: Differences in model squared Sharpe ratios}} \\
\midrule
           & {MACRO}  & {HKMNT}  & {LIQPS}   & {LIQAM}  & {VOLPS}  & {VOLAM}   \\ \midrule
{CAPMB}    & $0.007$ & $0.025$   & $-0.003$  & $0.021$  & $0.102$  & $0.109$   \\
           & [0.411]  & [0.539]  & [0.959]   & [0.662]  & [0.089]  &  [0.019]  \\

\end{tabular}

\vspace*{0.1in}
\setlength\tabcolsep{0.6mm}
\begin{tabular}{lccccc}
\multicolumn{6}{c}{Panel~D: Mimicking portfolio statistics for long-run consumption} \\ \midrule
 & \multicolumn{2}{c}{EJN basis assets} &  & \multicolumn{2}{c}{DMR basis assets} \\ \midrule
 & 2004:08-2016:12 & 1984:03-2019:12 &  & 2004:08-2016:12 & 1984:03-2019:12 \\ \midrule
Mean & $-$0.019 & 0.533 &  & 0.150 & 0.457 \\
$se_{\text{EJN}}$ & (0.196) & (0.061) &  & (0.213) & (0.075) \\
$se_{\text{DMR}}$ & (0.325) & (0.238) &  & (0.665) & (0.155) \\
 &  & \multicolumn{1}{l}{} & \multicolumn{1}{l}{} &  &  \\
Alpha  & $-$0.250 & 0.247 &  & $-$0.053 & 0.213 \\
$se_{\text{EJN}}$ & (0.216) & (0.058) & \multicolumn{1}{l}{} & (0.230) & (0.084) \\
$se_{\text{DMR}}$ & (0.316) & (0.223) & \multicolumn{1}{l}{} & (0.746) & (0.150) \\
 \bottomrule
\end{tabular}
\end{table}
\endgroup

\clearpage
\newgeometry{left=1.75cm, right=1.75cm, top=2cm, bottom=2.75cm}
\begin{adjustbox}{max width= 1\linewidth,center}
\begin{threeparttable}
\textbf{Table~5} \newline  \footnotesize{Risk premia and CSR R$^2$s for nontraded-factor models. \\ [0.04in]
\hspace*{0.1in}
The table presents the estimation results of six beta-pricing models with traded and nontraded factors. The models include the two-factor model of Bali, Subrahmanyam, and Wen  (2021, MACRO) with uncertainty risk ($UNC$), the two-factor model of  He, Kelly, and Manela (2017, HKMNT) with nontraded intermediary capital risk ($CPTL$), the six-factor liquidity risk models of Lin, Wang, and Wu (2011)  with the P\'{a}stor-Stambaugh ($PS$) and the Amihud ($AM$) illiquidity proxies (LIQPS and LIQAM, respectively), and the seven-factor systematic volatility ($VIX$) risk models of Chung, Wang, and Wu (2019) with the P\'{a}stor-Stambaugh and the Amihud illiquidity proxies (VOLPS and VOLAM, respectively).  The models are estimated using monthly excess returns on the 32 combination (combo) portfolios over the period 2004:08 to 2016:12 (149 months). We report parameter estimates $\hat{\gamma}$ (multiplied by 100) in Panel~A and $\hat{\lambda}$ (with $\hat{\lambda}_0$ multiplied by 100) in Panel~B, the  $t$-statistics under correctly specified models (\textit{t}-$\text{stat}_c$), and the model misspecification-robust \textit{t}-statistics (\textit{t}-$\text{stat}_m$) from Kan, Robotti, and Shanken (2013). In Panel~C, we report the OLS and GLS CSR R$^2$ for each model and the difference between the CSR R$^2$s of the models in row \textit{i} and column \textit{j}. $t$-statistics are in round brackets and $p$-values are in square brackets.}
\vspace{0.1in}
\label{tab:Table_Non_Traded_Risk_Premia}
\footnotesize
\setlength\tabcolsep{0.15mm}
\centering
\begin{tabular}{lccccccccccccccccccccccc}
\toprule
\multicolumn{24}{c}{Panel A: Price of beta risk (OLS)} \\ \midrule
 & \multicolumn{3}{c}{MACRO} &  & \multicolumn{3}{c}{HKMNT} &  & \multicolumn{3}{c}{LIQPS} &  & \multicolumn{3}{c}{LIQAM} &  & \multicolumn{3}{c}{VOLPS} &  & \multicolumn{3}{c}{VOLAM} \\ \midrule
 & $\hat{\gamma}_{0}$ & $\hat{\gamma}_{MKTB}$ & $\hat{\gamma}_{UNC}$ &  & $\hat{\gamma}_{0}$ & $\hat{\gamma}_{MKTS}$ & $\hat{\gamma}_{CPTL}$  &  & $\hat{\gamma}_{0}$ & $\hat{\gamma}_{MKTS}$ & $\hat{\gamma}_{PS}$ &  & $\hat{\gamma}_{0}$ & $\hat{\gamma}_{MKTS}$ & $\hat{\gamma}_{AM}$ &  & $\hat{\gamma}_{0}$ & $\hat{\gamma}_{MKTS}$ & $\hat{\gamma}_{VIX}$&  & $\hat{\gamma}_{0}$ & $\hat{\gamma}_{MKTS}$ & $\hat{\gamma}_{VIX}$ \\ \midrule
Estimate & 0.09 & 0.39 &   $-0.36$ & & 0.16 & 1.21 &  0.48 & & $0.17$ & 0.77 & $-0.64$ &  & 0.20 & 1.10 & $-0.33$ &  & 0.18 & 0.77 & $-0.40$ &  & 0.20 & 0.97 & $-0.86$  \\
\textit{t}-$\text{stat}_c$ & (0.90) & (2.09) & ($-1.11$) &  & (1.59) & (1.65) & (0.42) & & (2.31) & (1.30) & ($-1.60$) &  & (2.38) & (1.54) & ($-0.83$) &  & (2.17) & (1.00) & ($-0.42$) &  & (2.44) & (1.14) & ($-0.71$) \\
\textit{t}-$\text{stat}_m$ & (0.89) & (2.03) & ($-1.09$) &  & (1.73) & (1.61) & (0.39) & & (2.36) & (1.28) & ($-1.50$) &  & (2.45) & (1.44) & ($-0.79$) &  & (2.20) & (0.95) & ($-0.39$) &  & (2.52) & (1.13) & ($-0.61$) \\
R$^2$ & 0.911   &  & & & 0.891   & & & &  0.956   & & & & 0.948   & &  & & 0.956   & & &  & 0.953   & &    \\
      & [0.450] &  & & & [0.583] & & & &  [0.570] & & & & [0.545] & &  & & [0.428] & & &  & [0.396] & &   \\
 &  &  &  &  &  &  &  &  &  &  &  &  &  &  &  &  &  &  &  &  &  &  & \\
 \multicolumn{24}{c}{Panel B: Price of beta risk (GLS)} \\ \midrule
  & \multicolumn{3}{c}{MACRO} &  & \multicolumn{3}{c}{HKMNT} &  & \multicolumn{3}{c}{LIQPS} &  & \multicolumn{3}{c}{LIQAM} &  & \multicolumn{3}{c}{VOLPS} &  & \multicolumn{3}{c}{VOLAM} \\ \midrule
 & $\hat{\gamma}_{0}$ & $\hat{\gamma}_{MKTB}$ & $\hat{\gamma}_{UNC}$ &  & $\hat{\gamma}_{0}$ & $\hat{\gamma}_{MKTS}$ & $\hat{\gamma}_{CPTL}$  &  & $\hat{\gamma}_{0}$ & $\hat{\gamma}_{MKTS}$ & $\hat{\gamma}_{PS}$ &  & $\hat{\gamma}_{0}$ & $\hat{\gamma}_{MKTS}$ & $\hat{\gamma}_{AM}$ &  & $\hat{\gamma}_{0}$ & $\hat{\gamma}_{MKTS}$ & $\hat{\gamma}_{VIX}$&  & $\hat{\gamma}_{0}$ & $\hat{\gamma}_{MKTS}$ & $\hat{\gamma}_{VIX}$ \\ \midrule
Estimate & 0.01 & 0.46 &   $-0.15$ & & 0.03 & 0.07 &  0.73 & & $0.03$ & 0.24 & $-0.15$ &  & 0.02 & 0.19 & $0.34$ &  & 0.02 & 0.18 & $-0.50$ &  & 0.02 & 0.19 & $-0.40$  \\
\textit{t}-$\text{stat}_c$ & (0.60) & (2.48) & ($-0.48$) &  & (1.28) & (0.14) & (0.75) & & (1.12) & (0.47) & ($-0.46$) &  & (0.95) & (0.39) & ($1.25$) &  & (0.93) & (0.36) & ($-0.97$) &  & (0.90) & (0.38) & ($-0.77$) \\
\textit{t}-$\text{stat}_m$ & (0.57) & (2.52) & ($-0.46$) &  & (1.23) & (0.13) & (0.62) & & (1.10) & (0.44) & ($-0.35$) &  & (0.88) & (0.35) & ($1.13$) &  & (0.88) & (0.33) & ($-0.94$) &  & (0.82) & (0.34) & ($-0.70$) \\
R$^2$ & 0.098   &  & & & 0.012   & & & &  0.093   & & & & 0.099   & &  & & 0.104   & & &  & 0.100   & &    \\
      & [0.001] &  & & & [0.000] & & & &  [0.001] & & & & [0.000] & &  & & [0.001] & & &  & [0.000] & &   \\
 &  &  &  &  &  &  &  &  &  &  &  &  &  &  &  &  &  &  &  &  &  &  & \\
 \multicolumn{24}{c}{Panel C: Price of covariance risk (OLS)} \\ \midrule
  & \multicolumn{3}{c}{MACRO} &  & \multicolumn{3}{c}{HKMNT} &  & \multicolumn{3}{c}{LIQPS} &  & \multicolumn{3}{c}{LIQAM} &  & \multicolumn{3}{c}{VOLPS} &  & \multicolumn{3}{c}{VOLAM} \\ \midrule
 & $\hat{\lambda}_{0}$ & $\hat{\lambda}_{MKTB}$ & $\hat{\lambda}_{UNC}$ &  & $\hat{\lambda}_{0}$ & $\hat{\lambda}_{MKTS}$ & $\hat{\lambda}_{CPTL}$  &  & $\hat{\lambda}_{0}$ & $\hat{\lambda}_{MKTS}$ & $\hat{\lambda}_{PS}$ &  & $\hat{\lambda}_{0}$ & $\hat{\lambda}_{MKTS}$ & $\hat{\lambda}_{AM}$ &  & $\hat{\lambda}_{0}$ & $\hat{\lambda}_{MKTS}$ & $\hat{\lambda}_{VIX}$&  & $\hat{\lambda}_{0}$ & $\hat{\lambda}_{MKTS}$ & $\hat{\lambda}_{VIX}$ \\ \midrule
Estimate & 0.09 & 7.96 &   $-6.26$ & & 0.16 & 13.04 &  $-4.95$ & & $0.17$ & $-4.87$ & $-7.57$ &  & 0.20 & 0.05 & $-6.95$ &  & 0.18 & $-4.83$ & $0.05$ &  & 0.20 & $-4.56$ & $-9.76$  \\
\textit{t}-$\text{stat}_c$ & (0.90) & (1.32) & ($-0.56$) &  & (1.59) & (1.23) & ($-0.83$) & & (2.31) & ($-0.74$) & ($-1.72$) &  & (2.38) & (0.01) & ($-1.05$) &  & (2.17) & ($-0.36$) & ($0.00$) &  & (2.44) & ($-0.33$) & ($-0.51$) \\
\textit{t}-$\text{stat}_m$ & (0.89) & (1.23) & ($-0.54$) &  & (1.73) & (1.10) & ($-0.72$) & & (2.36) & ($-0.76$) & ($-1.71$) &  & (2.45) & (0.01) & ($-1.02$) &  & (2.20) & ($-0.35$) & ($0.00$) &  & (2.52) & ($-0.31$) & ($-0.44$) \\
 &  &  &  &  &  &  &  &  &  &  &  &  &  &  &  &  &  &  &  &  &  &  & \\
\multicolumn{24}{c}{Panel D: Price of covariance risk (GLS)} \\ \midrule
  & \multicolumn{3}{c}{MACRO} &  & \multicolumn{3}{c}{HKMNT} &  & \multicolumn{3}{c}{LIQPS} &  & \multicolumn{3}{c}{LIQAM} &  & \multicolumn{3}{c}{VOLPS} &  & \multicolumn{3}{c}{VOLAM} \\ \midrule
  & $\hat{\lambda}_{0}$ & $\hat{\lambda}_{MKTB}$ & $\hat{\lambda}_{UNC}$ &  & $\hat{\lambda}_{0}$ & $\hat{\lambda}_{MKTS}$ & $\hat{\lambda}_{CPTL}$  &  & $\hat{\lambda}_{0}$ & $\hat{\lambda}_{MKTS}$ & $\hat{\lambda}_{PS}$ &  & $\hat{\lambda}_{0}$ & $\hat{\lambda}_{MKTS}$ & $\hat{\lambda}_{AM}$ &  & $\hat{\lambda}_{0}$ & $\hat{\lambda}_{MKTS}$ & $\hat{\lambda}_{VIX}$&  & $\hat{\lambda}_{0}$ & $\hat{\lambda}_{MKTS}$ & $\hat{\lambda}_{VIX}$ \\ \midrule
Estimate & 0.01 & 13.92 &   $2.31$ & & 0.03 & $-3.55$ &  $3.22$ & & $0.03$ & $-3.48$ & $-2.45$ &  & 0.02 & $-3.05$ & $3.69$ &  & 0.02 & $-6.13$ & $-3.83$ &  & 0.02 & $-3.89$ & $-1.52$  \\
\textit{t}-$\text{stat}_c$ & (0.60) & (2.27) & ($0.23$) &  & (1.28) & ($-0.68$) & (0.86) & & (1.12) & ($-0.89$) & ($-0.79$) &  & (0.95) & ($-0.65$) & ($0.76$) &  & (0.93) & ($-1.49$) & ($-0.80$) &  & (0.90) & ($-0.79$) & ($-0.24$) \\
\textit{t}-$\text{stat}_m$ & (0.57) & (2.40) & ($0.22$) &  & (1.23) & ($-0.51$) & (0.64) & & (1.10) & ($-0.70$) & ($-0.59$) &  & (0.88) & ($-0.56$) & ($0.67$) &  & (0.88) & ($-1.10$) & ($-0.74$) &  & (0.82) & ($-0.61$) & ($-0.19$) \\
 &  &  &  &  &  &  &  &  &  &  &  &  &  &  &  &  &  &  &  &  &  &  & \\
\end{tabular}

\setlength\tabcolsep{1.10mm}
\begin{tabular}{lccccccccccccccccccccccc}
\multicolumn{24}{c}{Panel E: Differences in CSR R$^2$s} \\ \midrule
  & \multicolumn{6}{c}{OLS} & \multicolumn{11}{l}{} & \multicolumn{6}{c}{GLS} \\ \midrule
        & MACRO &  HKMNT &  LIQPS &  LIQAM &  VOLPS  & VOLAM & & & & & & & & & &  & & MACRO & HKMNT & LIQPS & LIQAM & VOLPS & VOLAM  \\ \midrule
CAPMB   & $-0.022$  & $-0.003$  & $-0.067$  & $-0.060$ & $-0.067$ & $-0.065$ & & & & & & & & & &  & &   $-0.001$  & 0.085 &  0.004 &  $-0.002$  & $-0.007$ & $-0.003$ \\
        & [0.589] &  [0.976]  & [0.372]  & [0.409] &  [0.371] &  [0.389]     & & & & & & & & & &  & &     [0.828] &  [0.194]  & [0.965] &  [0.979] &  [0.931]  & [0.972]\\ \bottomrule
\end{tabular}
\end{threeparttable}
\end{adjustbox}
\restoregeometry

\clearpage
\newgeometry{left=1.75cm, right=1.75cm, top=2cm, bottom=2.75cm}
\begin{adjustbox}{max width= 1\linewidth,center}
\begin{threeparttable}
\textbf{Table~6} \newline  \footnotesize{Fama-MacBeth estimates and \textit{t}-statistics of the zero-beta rate and risk premia for traded- and nontraded-factor models. \\ [0.04in]
\hspace*{0.1in}
The table presents the estimation results for six beta-pricing models with traded factors only and for six beta-pricing models with traded and nontraded factors. The traded-factor models include the CAPM with the corporate bond market factor (CAPMB), the equity CAPM with the stock market factor (CAPM), the four-factor model of Bai, Bali, and Wen (2019, BBW), the two-factor model of He, Kelly, and Manela (2017, HKM), the single-factor model of HKM (HKMSF), and the two-factor model with default and term structure factors of Fama and French (1993, DEFTERM). The nontraded-factor models include the two-factor model of Bali, Subrahmanyam, and Wen (2021, MACRO) with uncertainty risk ($UNC$), the two-factor model of  He, Kelly, and Manela (2017, HKMNT) with nontraded intermediary capital risk ($CPTL$), the six-factor liquidity risk models of Lin, Wang, and Wu (20111)  with the P\'{a}stor-Stambaugh ($PS$) and the Amihud ($AM$) illiquidity proxies (LIQPS and LIQAM, respectively), and the seven-factor systematic volatility ($VIX$) risk models of Chung, Wang, and Wu (2019) with the P\'{a}stor-Stambaugh and the Amihud illiquidity proxies (VOLPS and VOLAM, respectively). The models are estimated using cross-sectional Fama-MacBeth regressions of individual corporate bond excess returns on the post-ranking betas (or covariances). Post-ranking betas/covariances are estimated at the portfolio level by regressing
post-ranking portfolio excess returns on factor returns and are assigned to each bond in the portfolio, while pre-ranking factor betas/covariances are obtained from individual bond excess returns and factors over rolling 36-month windows with a minimum of 24 months required to include the coefficient estimate in the sample.  We report parameter estimates $\hat{\gamma}$ (multiplied by 100) in Panels~A and C and $\hat{\lambda}$ in Panels~B and D.  The sample period is 2004:08 to 2016:12 (149 months). After computing the rolling betas, the sample spans the period 2006:08 to 2016:12 (125 months). The Fama-MacBeth $t$-statistics (\textit{t}-$\text{stat}_{FM}$) with a 12-lag NW adjustment are in round brackets. We denote by Obs. the total number of observations and by Adj.~R$^2$ the average adjusted cross-sectional R$^2.$
}
\vspace{0.1in}
\label{tab:Table_FMB_Premia}
\footnotesize
\setlength\tabcolsep{0.15mm}
\centering
\begin{tabular}{lcccccccccccccccccc} \midrule
\multicolumn{19}{c}{Panel~A: Price of beta risk for traded-factor models} \\ \midrule
\multicolumn{1}{c}{} & \multicolumn{2}{c}{CAPMB} & \multicolumn{5}{c}{BBW} & \multicolumn{3}{c}{DEFTERM} & \multicolumn{2}{c}{CAPM} & \multicolumn{2}{c}{HKMSF} & \multicolumn{3}{c}{HKM} &  \\ \midrule
 & $\hat{\gamma}_{0}$ & $\hat{\gamma}_{MKTB}$ & $\hat{\gamma}_{0}$ & $\hat{\gamma}_{MKTB}$ & $\hat{\gamma}_{DRF}$ & $\hat{\gamma}_{CRF}$ & $\hat{\gamma}_{LRF}$ & $\hat{\gamma}_{0}$ & $\hat{\gamma}_{DEF}$ & $\hat{\gamma}_{TERM}$ & $\hat{\gamma}_{0}$ & $\hat{\gamma}_{MKTS}$ & $\hat{\gamma}_{0}$ & $\hat{\gamma}_{CPTLT}$ & $\hat{\gamma}_{0}$ & $\hat{\gamma}_{MKTS}$ & $\hat{\gamma}_{CPTLT}$ &  \\ \midrule
Estimate & 0.00 & 0.54 & 0.10 & 0.36 & 0.04 & 0.40 & 0.15 & 0.07 & 0.43 & 0.62 & 0.14 & 1.73 & 0.21 & 3.00 & $-$0.47 & 3.98 & $-$2.57 &  \\
\textit{t}-$\text{stat}_{FM}$ & (0.02) & (1.54) & (1.11) & (1.97) & (0.18) & (0.75) & (2.03) & (0.69) & (1.23) & (1.82) & (1.16) & (1.43) & (1.91) & (1.40) & ($-$1.01) & (1.47) & ($-$1.20) &  \\
Adj. R$^2$ & 0.058 &  & 0.111 &  &  &  &  & 0.079 &  &  & 0.069 &  & 0.062 &  & 0.034 &  &  &  \\
Obs.       & 321,280 &  & 321,280 &  &  &  &  & 321,280 &  &  & 321,280 &  & 321,280 &  & 321,280 &  &  &  \\
 &  &  &  &  &  &  &  &  &  &  &  &  &  &  &  &  &  &  \\
\multicolumn{19}{c}{Panel~B: Price of covariance risk for traded-factor models} \\ \midrule
\multicolumn{1}{c}{} & \multicolumn{2}{c}{CAPMB} & \multicolumn{5}{c}{BBW} & \multicolumn{3}{c}{DEFTERM} & \multicolumn{2}{c}{CAPM} & \multicolumn{2}{c}{HKMSF} & \multicolumn{3}{c}{HKM} &  \\ \midrule
 & $\hat{\lambda}_{0}$ & $\hat{\lambda}_{MKTB}$ & $\hat{\lambda}_{0}$ & $\hat{\lambda}_{MKTB}$ & $\hat{\lambda}_{DRF}$ & $\hat{\lambda}_{CRF}$ & $\hat{\lambda}_{LRF}$ & $\hat{\lambda}_{0}$ & $\hat{\lambda}_{DEF}$ & $\hat{\lambda}_{TERM}$ & $\hat{\lambda}_{0}$ & $\hat{\lambda}_{MKTS}$ & $\hat{\lambda}_{0}$ & $\hat{\lambda}_{CPTLT}$ & $\hat{\lambda}_{0}$ & $\hat{\lambda}_{MKTS}$ & $\hat{\lambda}_{CPTLT}$ &  \\ \midrule
Estimate & 0.04 & 12.25 & 0.16 & $-$0.64 & 1.08 & 4.37 & 6.49 & 0.30 & 9.23 & $-$2.81 & 0.15 & 8.48 & 0.23 & 5.04 & 0.15 & 5.39 & 2.18 &  \\
\textit{t}-$\text{stat}_{FM}$ & (0.37) & (1.51) & (1.85) & ($-$0.21) & (0.72) & (1.12) & (1.84) & (2.57) & (1.39) & ($-$0.71) & (1.31) & (1.42) & (2.01) & (1.38) & (1.28) & (1.48) & (1.02) &  \\
Adj. R$^2$ & 0.051 &  & 0.105 &  &  &  &  & 0.089 &  &  & 0.066 &  & 0.060 &  & 0.072 &  &  &  \\
Obs.       & 321,280 &  & 321,280 &  &  &  &  & 321,280 &  &  & 321,280 &  & 321,280 &  & 321,280 &  &  &  \\
 &  &  &  &  &  &  &  &  &  &  &  &  &  &  &  &  &  &  \\
\multicolumn{19}{c}{Panel~C: Price of beta risk for nontraded-factor models} \\ \midrule
 & \multicolumn{3}{c}{MACRO} & \multicolumn{3}{c}{HKMNT} & \multicolumn{3}{c}{LIQPS} & \multicolumn{3}{c}{LIQAM} & \multicolumn{3}{c}{VOLPS} & \multicolumn{3}{c}{VOLAM} \\ \midrule
 & $\hat{\gamma}_{0}$ & $\hat{\gamma}_{MKTB}$ & $\hat{\gamma}_{UNC}$ & $\hat{\gamma}_{0}$ & $\hat{\gamma}_{MKTS}$ & $\hat{\gamma}_{CPTL}$ & $\hat{\gamma}_{0}$ & $\hat{\gamma}_{MKTS}$ & $\hat{\gamma}_{PS}$ & $\hat{\gamma}_{0}$ & $\hat{\gamma}_{MKTS}$ & $\hat{\gamma}_{AM}$ & $\hat{\gamma}_{0}$ & $\hat{\gamma}_{MKTS}$ & $\hat{\gamma}_{VIX}$ & $\hat{\gamma}_{0}$ & $\hat{\gamma}_{MKTS}$ & $\hat{\gamma}_{VIX}$ \\ \midrule
Estimate & 0.04 & 0.44 & $-$0.80 & $-$0.21 & 2.82 & $-$2.53 & 0.14 & 1.31 & $-$2.92 & $-$0.05 & 2.68 & $-$0.71 & 0.06 & 2.76 & 0.18 & 0.44 & 1.80 & 0.04 \\
\textit{t}-$\text{stat}_{FM}$ & (0.17) & (1.24) & ($-$1.31) & ($-$0.79) & (1.52) & ($-$1.19) & (1.15) & (1.34) & ($-$1.25) & ($-$0.33) & (1.73) & ($-$0.79) & (0.48) & (1.71) & (0.28) & (1.76) & (1.59) & (0.05) \\
Adj. R$^2$ & 0.051 &  & & 0.056   &  &    & 0.099 & &    & 0.103 &  & & 0.088 &  & & 0.087 &  &   \\
Obs.       & 321,280 &  & & 321,280 &  &    & 321,280 &  & &   321,280 & &  & 321,280 &  & & 321,280 &  &    \\
 &  &  &  &  &  &  &  &  &  &  &  &  &  &  &  &  &  &  \\
\multicolumn{19}{c}{Panel~D: Price of covariance risk for nontraded-factor models} \\ \midrule
 & \multicolumn{3}{c}{MACRO} & \multicolumn{3}{c}{HKMNT} & \multicolumn{3}{c}{LIQPS} & \multicolumn{3}{c}{LIQAM} & \multicolumn{3}{c}{VOLPS} & \multicolumn{3}{c}{VOLAM} \\ \midrule
 & $\hat{\lambda}_{0}$ & $\hat{\lambda}_{MKTB}$ & $\hat{\lambda}_{UNC}$ & $\hat{\lambda}_{0}$ & $\hat{\lambda}_{MKTS}$ & $\hat{\lambda}_{CPTL}$ & $\hat{\lambda}_{0}$ & $\hat{\lambda}_{MKTS}$ & $\hat{\lambda}_{PS}$ & $\hat{\lambda}_{0}$ & $\hat{\lambda}_{MKTS}$ & $\hat{\lambda}_{AM}$ & $\hat{\lambda}_{0}$ & $\hat{\lambda}_{MKTS}$ & $\hat{\lambda}_{VIX}$ & $\hat{\lambda}_{0}$ & $\hat{\lambda}_{MKTS}$ & $\hat{\lambda}_{VIX}$ \\ \midrule
Estimate & $-$0.08 & 3.45 & $-$20.38 & 0.03 & 4.79 & 4.75 & 0.03 & 3.77 & 9.98 & 0.33 & 4.01 & $-$7.77 & 0.02 & 3.93 & $-$0.04 & 0.26 & 3.87 & $-$1.36 \\
\textit{t}-$\text{stat}_{FM}$ & ($-$0.36) & (0.75) & ($-$1.52) & (0.18) & (1.22) & (1.49) & (0.21) & (1.40) & (1.96) & (2.67) & (1.41) & ($-$1.17) & (0.19) & (1.44) & ($-$0.02) & (1.68) & (1.39) & ($-$0.69) \\
Adj. R$^2$ & 0.071 &  & & 0.074 &     & & 0.114 &  &    & 0.113 & & & 0.116 &  & & 0.115 &   &  \\
Obs.       & 321,280 &  & & 321,280 &  &    & 321,280 &  & &   321,280 & &  & 321,280 &  & & 321,280 &  &    \\ \bottomrule
\end{tabular}
\end{threeparttable}
\end{adjustbox}
\restoregeometry

\setcounter{table}{0}
\renewcommand{\thetable}{A\arabic{table}}

\clearpage
\begin{table}[h!]
\fontsize{11pt}{19pt}\selectfont
\textbf{Table~A1} \newline  \footnotesize{Number of bonds and issuers. \\[0.04in] \hspace*{0.1in}
Panel~A reports the total number of bond-month return observations for which there are also valid observations on three bond characteristics (credit rating, time-to-maturity, and amount outstanding), the total number of bonds and firms, and the average number of bonds and firms in any given month $t$ based on TRACE and Bai, Bali, and Wen (2019, BBW). In Panel~B, we vary the number of days a bond must trade toward the end or beginning of the month from $n = 1$ (the bond must trade on the last business day of the month) to $n>10$ (the bond can trade on any day of the month). The sample period is 2002:07 to 2016:12 (174 months).}
\vspace{0.1in} \par
\label{tab:Table_Bond_Firms}
\begin{tabular*}{\textwidth}{@{\extracolsep{\fill}}lccccc}
\toprule
 \multicolumn{6}{c}{Panel   A: Number of bonds and issuers in TRACE} \\ \midrule
 & \multicolumn{1}{l}{Total Obs.} & Total bonds & Total firms & Average bonds in month & Average firms in month \\ \midrule
TRACE & 861,524 & 31,348 & 3,792 & 4,951 & 1,308 \\
BBW & 1,243,543 & 38,957 & 4,079 & 7,147 & - \\
 & \multicolumn{1}{l}{} &  &  &  &  \\
\multicolumn{6}{c}{Panel B: Number of bonds and issuers in TRACE across $n$} \\ \midrule
\multicolumn{1}{c}{} & \multicolumn{1}{l}{Total Obs.} & Total bonds & Total firms & Average bonds in month & Average firms in month \\ \midrule
TRACE$_{n=1}$ & 518,566 & 25,290 & 3,520 & 2,997 & 915 \\
TRACE$_{n=3}$ & 673,626 & 28,039 & 3,670 & 3,871 & 1,106 \\
TRACE$_{n=5}$ & 861,524 & 31,348 & 3,792 & 4,951 & 1,308 \\
TRACE$_{n=7}$ & 976,668 & 31,802 & 3,806 & 5,613 & 1,418 \\
TRACE$_{n=10}$ & 1,104,759 & 32,195 & 3,812 & 6,349 & 1,524 \\
TRACE$_{n>10}$ & 1,254,736 & 32,507 & 3,821 & 7,211 & 1,624\\ \bottomrule
\end{tabular*}
\end{table}

\clearpage
\begingroup
\fontsize{11pt}{19pt}\selectfont
\begin{table}[h!]
\label{tab:Table_Descriptives_TRACE}
\textbf{Table~A2} \newline  \footnotesize{Descriptive statistics. \\[0.04in] \hspace*{0.1in} Panel~A reports the number of bond-month observations ($N$), the time-series average of the cross-sectional mean (Mean), median (Median), standard deviation (SD), and percentiles (Percentiles) of monthly corporate bond returns (Return, \%) and bond characteristics including credit rating (Rating), time-to-maturity (Maturity, year), and amount outstanding
(Size, \$ million). Ratings are converted to numerical scores, where 1 refers to a AAA-rated bond and 22 refers to a D rating where the bond issue is technically in default. A higher score implies a higher level of credit risk. Numerical ratings of 10 or below (BBB$-$ or better) are classified as investment grade, and ratings of 11 or higher (BB$+$ or worse) are labeled speculative (junk) grade. Downside risk is the 5\% VaR of the corporate bond return (VaR), defined as the second lowest monthly return observation over the past 36 months scaled by $-1$ (using a minimum of 24-months of data). Illiquidity ($ILLIQ$) is computed as in Bao, Pan, and Wang (2011). Panel~B reports the time-series average of the cross-sectional correlations. Panels~A and B are based on the sample period 2002:07 to 2016:12 (174 months) for all variables except for VaR, which spans the sample period 2004:06 to 2016:12 (151 months).}
\vspace{0.1in} \par
\footnotesize
\begin{tabular*}{\textwidth}{@{\extracolsep{\fill}}lcccccccccc}
\toprule
\multicolumn{11}{c}{Panel~A: Average cross-sectional statistics} \\ \midrule
 & \multicolumn{1}{l}{} & \multicolumn{1}{l}{} & \multicolumn{1}{l}{} & \multicolumn{1}{l}{} & \multicolumn{6}{c}{Percentiles} \\ \cmidrule(lr){6-11}
 & $N$ & Mean & Median & SD & 1 & 5 & 25 & 75 & 95 & 99 \\ \midrule
Return  & 861,524 & 0.72 & 0.52 & 4.19 & $-$8.98 & $-$4.04 & $-$0.66 & 1.86 & 6.08 & 12.60 \\
Rating & 861,524 & 8.67 & 8.06 & 4.11 & 1.32 & 2.18 & 5.84 & 10.66 & 16.34 & 19.94 \\
Maturity & 861,524 & 9.39 & 6.47 & 8.54 & 1.11 & 1.54 & 3.63 & 11.66 & 26.99 & 29.89 \\
Size & 861,524 & 492 & 346 & 559 & 3 & 10 & 152 & 621 & 1,523 & 2,706 \\
VaR & 417,699 & 5.27 & 3.69 & 5.05 & 0.68 & 1.07 & 2.21 & 6.34 & 15.05 & 25.35 \\
$ILLIQ$ & 711,128 & 1.70 & 0.33 & 4.71 & $-$1.08 & $-$0.23 & 0.06 & 1.39 & 8.50 & 20.22 \\ \bottomrule
\end{tabular*}

\vspace*{2 mm}
\begin{tabular*}{\textwidth}{@{\extracolsep{\fill}}lcccccc}
\multicolumn{7}{c}{Panel~B: Average cross-sectional correlations} \\ \midrule
 & Return & Rating & Maturity & Size & VaR & $ILLIQ$ \\ \midrule
Return & 1 & 0.07 & 0.02 & $-$0.02 & 0.07 & 0.01 \\
Rating &  & 1 & $-$0.13 & $-$0.07 & 0.50 & 0.18 \\
Maturity &  &  & 1 & $-$0.02 & 0.20 & 0.13 \\
Size &  &  &  & 1 & $-$0.11 & $-$0.14 \\
VaR &  &  &  &  & 1 & 0.33 \\
$ILLIQ$ &  &  &  &  &  & 1 \\ \bottomrule
\end{tabular*}
\end{table}
\endgroup

\begin{figure}[p]
\includegraphics[width=1\textwidth]{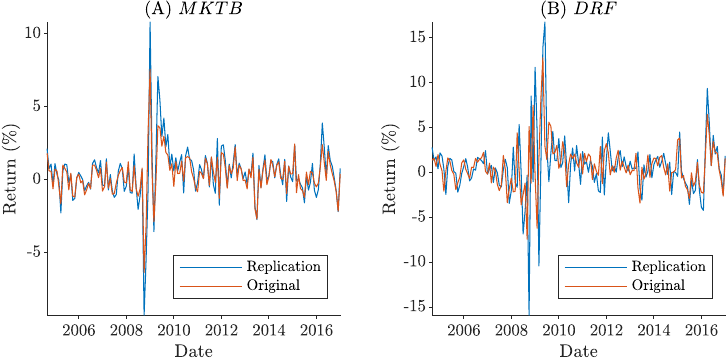} \\ \vspace*{0.1in}

\includegraphics[width=1\textwidth]{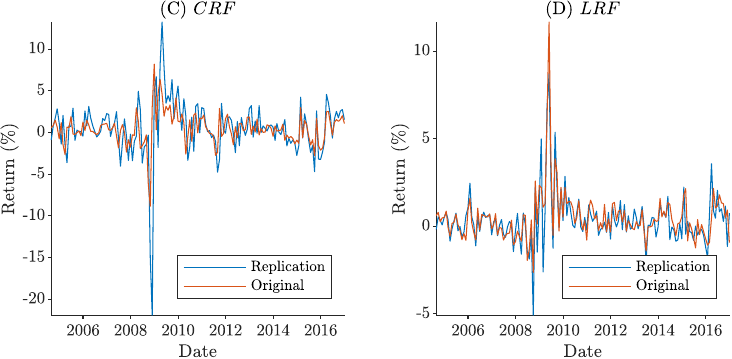}\\

\textbf{Fig.~1.}%
~Replicated and original BBW factors. The figure plots the replicated and original four factors of Bai, Bali, and Wen (2019, BBW). The factors include the value-weighted bond market factor ($MKTB$), the downside risk factor ($DRF$), the credit risk factor ($CRF$), and the liquidity risk factor ($LRF$). The sample period for all factors is 2004:08 to 2016:12 (149 months). The returns are monthly and presented in percent.
\end{figure}

\begin{figure}[p]
\includegraphics[width=1\textwidth]{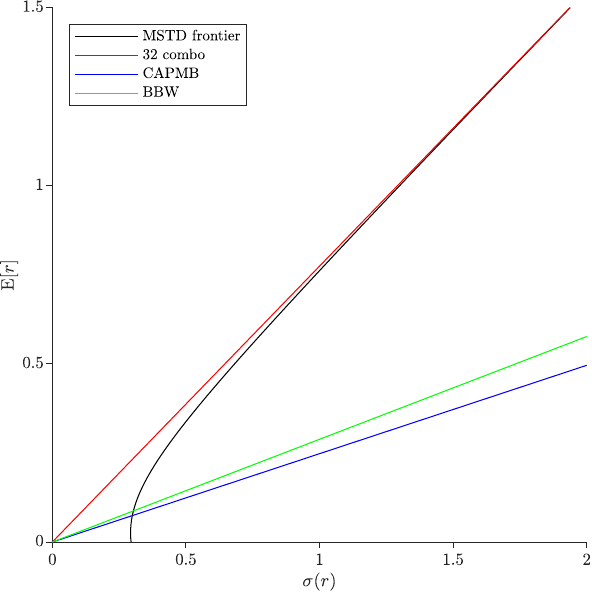} \\

 \textbf{Fig.~2.}%
~Mean-standard deviation frontier and maximum Sharpe ratios. The figure plots the mean-standard deviation frontier (MSTD frontier, black line) for the 32 combination (combo) portfolios, which comprise 10 portfolios sorted on credit spreads, 5 portfolios sorted on maturity, 5 portfolios sorted on bond rating, and the Fama-French 12 industry portfolios.
The slope of the red line represents
the maximum Sharpe ratio that can be achieved by investing in the 32 combination (combo) portfolios.
Similarly, the slope of the green line ({BBW}) represents the maximum Sharpe ratio
from optimally combining the four factors of Bai, Bali, and Wen (2019, BBW). The slope of the blue line ({CAPMB}) is the Sharpe ratio of $MKTB,$ the bond market factor. The sample period  is 2004:08 to 2016:12 (149 months). The mean return ($E[r],$ $y$-axis) and return standard deviation ($\sigma(r),$ $x$-axis) are monthly and presented in percent.
\end{figure}

\clearpage

\begingroup

\begin{thebibliography}{00}

\bibitem{} Amihud, Y., 2002. Illiquidity and stock returns: Cross-section and time-series effects.
Journal of Financial Markets 5, 31--56.

\bibitem{} Bai, J., Bali, T.~G., Wen, Q., 2019. Common risk factors in the cross-section of corporate bond returns.
Journal of Financial Economics 131, 619--642.

\bibitem{} Bai, J., Bali, T.~G., Wen, Q., 2023. Replication of BBW factors with WRDS data.
Unpublished working paper. Georgetwon University.

\bibitem{} Bali, T.~G., Subrahmanyam, A., Wen, Q., 2021. The macroeconomic uncertainty premium in the corporate bond market.
Journal of Financial and Quantitative Analysis 56, 1653--1678.


\bibitem{} Bandi, F.~M., Chaudhuri, S.~E., Lo, A.~W., Tamoni, A., 2021. Spectral factor models.
Journal of Financial Economics 142, 214--238.


\bibitem{} Bao, J., Pan, J., Wang, J., 2011. The illiquidity of corporate bonds.
Journal of Finance 66, 911--946.

\bibitem{} Barillas, F., Kan, R., Robotti, C., Shanken, J., 2020.
Model comparison with Sharpe ratios. Journal of Financial and Quantitative
Analysis 55, 1840--1874.


\bibitem{} Binsbergen, J.~H.~van, Nozawa, Y., Schwert, M., 2023.
Duration-based valuation of corporate bonds. Unpublished working paper. Jacobs Levy Equity Management Center for Quantitative Financial Research.

\bibitem{} Chung, K.~H., Wang, J., Wu, C., 2019.
Volatility and the cross-section of corporate bond returns. Journal of Financial Economics 133, 397--417.


\bibitem{} Cochrane, J.~H., 2005. Asset Pricing. Princeton University Press,
Princeton and Oxford.



\bibitem{} Detzel, A., Novy-Marx, R., Velikov, M., 2023. Model comparison with transaction costs.
Journal of Finance, forthcoming.


\bibitem{} Dick-Nielsen, J., 2014. How to clean Enhanced TRACE data.
Capital Markets: Market Microstructure eJournal.

\bibitem{} Dick-Nielsen, J., Rossi, M., 2019. The cost of immediacy for corporate bonds.
Review of Financial Studies 32, 1--41.


\bibitem{} Elkamhi, R., Jo, C., Nozawa, Y., 2023. A one-factor model of corporate bond premia. Management Science, forthcoming.


\bibitem{} Fama, E.~F., French, K.~R., 1992. The cross-section of expected stock returns.
Journal of Finance 47, 427--465.


\bibitem{} Fama, E.~F., French, K.~R., 1993. Common risk factors in the
returns on stocks and bonds. Journal of Financial Economics 33, 3--56.


\bibitem{} Fama, E.~F., MacBeth, J.~D., 1973. Risk, return, and equilibrium:
Empirical tests. Journal of Political Economy 71, 607--636.


\bibitem{} Gagliardini, P., Ossola, E., Scaillet, O., 2016. Time-varying risk premium in large cross-sectional equity data sets. Econometrica 84, 985--1046.


\bibitem{} Gibbons, M.~R., Ross, S.~A., Shanken, J., 1989. A test of the
efficiency of a given portfolio. Econometrica 57, 1121--1152.

\bibitem{} Goldberg, J., Nozawa, Y., 2021. Liquidity supply in the corporate bond market. Journal of Finance 76, 755--796.

\bibitem{} Gospodinov, N., Robotti, C., 2021a. Capital share risk in U.S. asset pricing:
A reappraisal. Journal of Finance, Replications and Comments.


\bibitem{} Gospodinov, N., Robotti, C., 2021b. Common pricing across asset classes:
Empirical evidence revisited. Journal of Financial Economics 140, 292--324.

\bibitem{} He, Z., Kelly, B., Manela, A., 2017. Intermediary asset pricing:
New evidence from many asset classes. Journal of Financial Economics 126,
1--35.

\bibitem{} Jurado, K., Ludvigson, S.~C., Ng, S., 2015. Measuring uncertainty. American Economic Review 105, 1177--1216.

\bibitem{} Kan, R., Robotti, C., 2012. Evaluation of asset pricing models
using two-pass cross-sectional regressions. In: Duan, J.-C., Gentle, J.~E.,
Haerdle, W. (Eds.), Handbook of Computational Finance. Springer, New York City.

\bibitem{} Kan, R., Robotti, C., Shanken, J., 2013. Pricing model
performance and the two-pass cross-sectional regression methodology. Journal
of Finance 68, 2617--2649.


\bibitem{} Kandel, S., Stambaugh, R.~F., 1995. Portfolio inefficiency and
the cross-section of expected returns. Journal of Finance 50, 157--184.

\bibitem{} Kleibergen, F., Zhan, Z., 2015. Unexplained factors and their
effects on second pass R-squared's. Journal of Econometrics 189, 101--116.

\bibitem{} Kleibergen, F., Zhan, Z., 2023. Identification robust inference for risk premia in short panels.
Unpublished working paper. University of Amsterdam.

\bibitem{} Kroencke, T.~A., Thimme, J., 2023. A skeptical appraisal of robust asset pricing tests.
Unpublished working paper. University of Neuchatel.


\bibitem{} Lewellen, J.~W., Nagel, S., Shanken, J., 2010. A skeptical
appraisal of asset pricing tests. Journal of Financial Economics 96,
175--194.

\bibitem{} Lin, H., Wang, J., Wu, C., 2011. Liquidity risk and expected corporate bond returns.
Journal of Financial Economics 99,
628--650.


\bibitem{} Neuhierl, A., Varneskov, R.~T., 2021. Frequency dependent risk. Journal of Financial Economics 140, 644--675.


\bibitem{} Newey, W.~K., West, K.~D., 1987. A simple, positive semi-definite, heteroskedasticity and autocorrelation consistent covariance matrix. Econometrica 55, 703--708.


\bibitem{} Nozawa, Y., 2017. What drives the cross-section of credit spreads?: A variance decomposition
approach. Journal of Finance 72, 2045--2072.


\bibitem{} P\'{a}stor, L., Stambaugh, R.~F., 2003. Liquidity risk and expected stock returns.
Journal of Political Economy 111,
642--685.



\end{thebibliography}
\endgroup
\end{document}